\documentclass[a4paper,UKenglish,cleveref,thm-restate]{lipics-v2021}

\newcommand{\Set}{\textnormal{Set}}

\newcommand{\C}{\mathscr{C}}
\newcommand{\EM}{{\C}^T}

\newcommand{\Pow}{\mathcal{P}}

\usepackage{tikz}
\usepackage{tikz-cd}
\usetikzlibrary{automata, positioning, arrows}
\usepackage{float}
\usepackage{enumitem}
\usepackage{caption}
\usepackage{subcaption}
\usepackage{wrapfig}
\usepackage{adjustbox}
\usepackage{stmaryrd}

\title{Generators and Bases for Monadic Closures} %TODO Please add

\author{Stefan Zetzsche\footnote{This paper is the result of work done prior to the author's affiliation with Amazon Web Services.}}{Amazon Web Services \and University College London \and \url{https://zetzsche.st}}{stefanzetzsche@gmail.com}{}{Prior to their affiliation with Amazon Web Services, supported by GCHQ via the VeTSS grant \emph{Automated Black-Box Verification of Networking Systems} (4207703/RFA 15845) and by the ERC via the Consolidator Grant \emph{AutoProbe} (101002697).}

\author{Alexandra Silva}{Cornell University \and University College London\and \url{https://alexandrasilva.org}}{alexandra.silva@gmail.com}{}{Supported by the ERC via the Consolidator Grant \emph{AutoProbe} (101002697) and by a Royal Society Wolfson Fellowship.}
\author{Matteo Sammartino}{Royal Holloway, University of London \and University College London \and \url{https://matteosammartino.com}}{Matteo.Sammartino@rhul.ac.uk}{}{Supported by the EPSRC Standard Grant \emph{CLeVer} (EP/S028641/1).}

\authorrunning{S. Zetzsche, A. Silva, and M. Sammartino} %TODO mandatory. First: Use abbreviated first/middle names. Second (only in severe cases): Use first author plus 'et al.'

\Copyright{Stefan Zetzsche and Alexandra Silva and Matteo Sammartino} %TODO mandatory, please use full first names. LIPIcs license is "CC-BY";  http://creativecommons.org/licenses/by/3.0/


 %TODO mandatory; please add comma-separated list of keywords

\category{} %optional, e.g. invited paper

\relatedversion{} %optional, e.g. full version hosted on arXiv, HAL, or other respository/website
\nolinenumbers %uncomment to disable line numbering

\EventEditors{Paolo Baldan and Valeria de Paiva}
\EventNoEds{2}
\EventLongTitle{10th Conference on Algebra and Coalgebra in Computer Science (CALCO 2023)}
\EventShortTitle{CALCO 2023}
\EventAcronym{CALCO}
\EventYear{2023}
\EventDate{June 19--21, 2023}
\EventLocation{Indiana University Bloomington, IN, USA}
\EventLogo{}
\SeriesVolume{270}
\ArticleNo{11}
\begin{document}

\maketitle

\begin{abstract}
  It is well-known that every regular language admits a unique minimal deterministic acceptor. Establishing an analogous result for non-deterministic acceptors is significantly more difficult, but nonetheless of great practical importance. To tackle this issue, a number of sub-classes of non-deterministic automata have been identified, all admitting canonical minimal representatives. In previous work, we have shown that such representatives can be recovered categorically in two steps. First, one constructs the minimal bialgebra accepting a given regular language, by closing the minimal coalgebra with additional algebraic structure over a monad. Second, one identifies canonical generators for the algebraic part of the bialgebra, to derive an equivalent coalgebra with side effects in a monad. In this paper, we further develop the general theory underlying these two steps. On the one hand, we show that deriving a minimal bialgebra from a minimal coalgebra can be realized by applying a monad on an appropriate category of subobjects. On the other hand, we explore the abstract theory of generators and bases for algebras over a monad.
\end{abstract}

\section{Introduction}
\label{sec:introduction}

The existence of a unique minimal \emph{deterministic} finite automaton is an important property of regular languages \cite{nerode1958linear}. Establishing a similar result for \emph{non-deterministic} finite automata is of great importance, as non-deterministic automata can be exponentially more succinct than deterministic ones, but turns out to be surprisingly difficult. The main problem is that a regular language can be accepted by several size-minimal NFAs that are not isomorphic. An example illustrating the situation is displayed in \Cref{example_nonisomorphic_nfa_canonical_automata_chapter}.

To tackle the issue, a number of sub-classes of non-deterministic automata admitting canonical representatives have been identified \cite{BrzozowskiT14, denis2001residual, VuilleminG210, MyersAMU15}. One such example is the \emph{canonical residual finite state automaton} (short \emph{canonical RFSA}, also known as \emph{jiromaton}), which is minimal among non-deterministic automata accepting joins of residual languages \cite{denis2001residual}. In previous work \cite{zetzsche2021}, we have presented a categorical framework that unifies constructions and correctness proofs of canonical non-deterministic automata and unveils new ones.

The framework adopts the well-known representation of automata as coalgebras \cite{jacobs2017introduction, rutten2019method, rutten2000universal} and side-effects like non-determinism as monads \cite{moggi1988computational, moggi1990abstract, moggi1991notions}. For instance, an NFA (without initial states) is represented as a coalgebra $( X,k ) $ with side-effects in the powerset monad $( \Pow, \{-\}, \mu )$, where $X$ is the set of states,  
$
	k \colon X \to 2 \times \Pow(X)^A
$
combines the function classifying each state as accepting or rejecting with the function giving the set of next states for each input, $\{-\}$ creates singleton sets, and $\mu$ takes the union of a set of sets. 

\begin{figure*}[t]
\centering
\begin{subfigure}{0.45 \textwidth}
\tiny
\centering
	\begin{tikzpicture}[node distance=4.2em]
	\node[state, initial, initial text=, minimum size=1.9em] (q1) {};
	\node[state, right of=q1, above of=q1, minimum size=1.9em] (q2) {};
		\node[state, below of=q2, minimum size=1.9em] (q3) {};
		\node[state, below of=q3, minimum size=1.9em] (q4) {};
		\node[state, right of=q3, minimum size=1.9em, accepting] (q5) {};
	    \path[->]
	(q1) edge[above] node{$a$} (q2)
		(q1) edge[above] node{$b$} (q3)
		(q1) edge[above] node{$c$} (q4)
		(q2) edge[right] node{$b,c$} (q5)
		(q3) edge[above] node{$a,c$} (q5)
		(q4) edge[right] node{$a,b$} (q5)
        ;
\end{tikzpicture}\qquad
\begin{tikzpicture}[node distance=4.2em]
	\node[state, initial, initial text=, minimum size=1.9em] (q1) {};
	\node[state, right of=q1, above of=q1, minimum size=1.9em] (q2) {};
		\node[state, below of=q2, minimum size=1.9em] (q3) {};
		\node[state, below of=q3, minimum size=1.9em] (q4) {};
		\node[state, right of=q3, minimum size=1.9em, accepting] (q5) {};
	    \path[->]
	(q1) edge[left] node{$b,c$} (q2)
		(q1) edge[above] node{$a,c$} (q3)
		(q1) edge[below] node{$a,b$} (q4)
		(q2) edge[right] node{$a$} (q5)
		(q3) edge[above] node{$b$} (q5)
		(q4) edge[right] node{$c$} (q5)
        ;
\end{tikzpicture}
\caption{Two non-isomorphic size-minimal NFA accepting $\lbrace ab,ac,$ $ba,bc,ca,cb \rbrace \subseteq \lbrace a, b, c \rbrace^*$ \cite{arnold1992note}.}
\label{example_nonisomorphic_nfa_canonical_automata_chapter}
\end{subfigure}	
\qquad
\begin{subfigure}{0.45 \textwidth}
\scriptsize
\centering
	\begin{tikzcd}[ ampersand replacement=\&, row sep = 0.75em]
		X \ar{d}[swap]{k} \ar{r}{\{-\}} \&
			\mathcal{P}(X) \ar{dl}{k^\sharp} \ar[dashed]{r}{\textnormal{obs}} \&
			2^{A^*} \ar{d}{\langle \varepsilon, \delta \rangle} \\
		2 \times \mathcal{P}(X)^A \ar[dashed]{rr}[below]{2 \times \textnormal{obs}^A} \&
			\&
			2 \times (2^{A^*})^A
	\end{tikzcd}
	\qquad
			\begin{tikzcd}[ampersand replacement=\&, row sep = 0.75em]
			X \ar{d}[swap]{k} \ar{r}{\eta} \&
			TX \ar{dl}{k^\sharp} \ar[dashed]{r}{\textnormal{obs}} \&
				\Omega \ar{d}{\omega} \\
			FTX \ar[dashed]{rr}[below]{F\textnormal{obs}} \&
				\&
				F\Omega
		\end{tikzcd}.
\caption{Classical and generalised determinisation of automata with side-effects in a monad \cite{silva2010generalizing}.}
\label{gen-det-diagrams}
\end{subfigure}
\caption{Non-isomorphic NFAs and generalised determinisation.}
\label{}
\end{figure*}

To derive canonical non-deterministic acceptors, the framework suggests a procedure that is closely related to the so-called \emph{powerset construction}. As depicted at the top of \Cref{gen-det-diagrams}, the latter converts a non-deterministic finite automaton $(X, k)$ into an equivalent deterministic finite automaton $(\mathcal{P}X, k^{\sharp})$, where $k^\sharp$ is obtained by lifting $k$ to the subsets of $X$, the tuple $\langle \varepsilon, \delta \rangle$ is the automaton of languages, and the morphism $\textnormal{obs}$ assigns language semantics to each set of states. As seen at the bottom of \Cref{gen-det-diagrams}, the construction can be generalised by replacing the functor $2 \times (-)^A$ with any (suitable) functor $F$ describing the automaton structure, and $\mathcal{P}$ with a monad $T$ describing the automaton side-effects, to transform a coalgebra $k: X \rightarrow FTX$ with side-effects in $T$ into an equivalent coalgebra $k^{\sharp}: TX \rightarrow FTX$  \cite{silva2010generalizing}. Under this perspective, $\Omega \xrightarrow{\omega} F \Omega$ is the so-called \emph{final coalgebra}, providing a semantic universe that generalises the automaton of languages. The deterministic automata resulting from such determinisation constructions have 
\emph{additional algebraic structure}: the state space $\Pow(X)$ defines a free complete join-semilattice (CSL) over $X$ and $k^\sharp$ is a CSL homomorphism. More generally, $TX$ defines a (free) algebra for the monad $T$, and $k^\sharp$ is a $T$-algebra homomorphism, thus constituting a so-called \emph{bialgebra} over a \emph{distributive law} relating $F$ and $T$ \cite{beck1969distributive, Street2009}. 

Using the powerset construction, a canonical succinct acceptor for a regular language $L \subseteq A^*$ over an alphabet $A$ can be obtained in two steps:
\begin{enumerate}
	\item One constructs the minimal
%\footnote{Minimal in the sense that every state is reachable by an element of $A^*$ and no two different states observe the same language.} 
(pointed) coalgebra $\textnormal{M}_L$ for the functor $F = 2 \times (-)^{A}$ accepting $L$. For the case $A = \lbrace a, b \rbrace$ and $L = (a+b)^*a$, the coalgebra $\textnormal{M}_{L}$ is depicted in \Cref{m(l)}. Generally, it can be obtained via the Myhill-Nerode construction \cite{nerode1958linear}. One then equips the former with additional algebraic structure in a monad $T$ (which is related to $F$ via a typically canonically\footref{canonicaldistributivelawfootnote} induced distributive law). This can be done by applying the generalised determinisation procedure to $\textnormal{M}_L$, when seen as coalgebra with trivial side-effects in $T$. By identifying semantically equivalent states one consequently derives the minimal
%\footnote{Minimal in the sense that every state is reachable by an element of $T(A^*)$ and no two different states observe the same language.} 
(pointed) bialgebra for $L$. If $T = \mathcal{P}$ is the powerset monad, the minimal bialgebra for the language $L = (a+b)^*a$ is depicted in \Cref{overlineml}.
	\item One exploits the algebraic structure underlying the minimal bialgebra for $L$ to  ``reverse'' the generalised determinisation procedure. That is, one identifies a minimal set of \emph{generators} that spans the full algebraic structure, to derive an equivalent succinct automaton with side-effects in $T$. For example, by choosing the \emph{join-irreducibles}\footnote{A \emph{join-irreducible} is a non-zero element $a$ satisfying, for all $y,z \in L$ with $a=y \vee z$, that $a = y $ or $a = z$.} for the CSL underlying the minimal bialgebra in \Cref{overlineml} as generators (in this case, the join-irreducibles are given by all non-zero states), one recovers the canonical acceptor in \Cref{jiromaton}.
\end{enumerate}

 In this paper, we further develop the general theory underlying these two steps. 
First,  we generalise the closure of a subset of an algebraic structure as a functor between categories of subobjects relative to a factorisation system. We then equip the functor with the structure of a monad.
%  (\Cref{inducedmonad})
	 We investigate the closure of a particular subclass of subobjects: the ones that arise from the image of a morphism. 
%	 (\Cref{closureofimage})
	 We show that deriving a minimal bialgebra from a minimal coalgebra can be realized by applying the monad to a subobject in this class. 
%	 (\Cref{closure_of_coalg})
%	
%	 
 Second, we define a category of algebras with generators, 
%(\Cref{generatorcategory})
 which is in adjunction with the category of Eilenberg-Moore algebras,
% (\Cref{galgfreeforgetful})
 and, under certain assumptions, monoidal.
%  (\Cref{monoidalproduct})
We generalise the matrix representation theory of vector spaces 
%(\Cref{representationtheorysec}) 
and discuss bases for bialgebras.
%(\Cref{basesforbialgebrasec})
We compare our ideas with an approach that generalises bases as coalgebras \cite{jacobs2011bases}.
% (\Cref{basesascoaglebrassec})
 We find that a basis in our sense induces a basis in the sense of \cite{jacobs2011bases},
%  (\Cref{impliesjacobsbasis})
  and identify assumptions under which the reverse is true, too.
	We characterise generators for finitary varieties in the sense of universal algebra 
%	(\Cref{sigmatermlemma}) 
	and relate our work to the theory of locally finitely presentable categories.
%	 (\Cref{finitarymonadlemma})

\begin{figure}[t]
	\centering
\begin{subfigure}[b]{0.26 \textwidth}
\centering
\tiny
							\adjustbox{valign=m}{
	\begin{tikzpicture}[node distance=6em]
	\node[state, initial, initial text=] (x) {$x$};	
		\node[state, right of=x, accepting] (y) {$y$};
	    \path[->]
	(x) edge[loop above] node{$b$} (x)
	(y) edge[loop right] node{$a$} (y)
	(x) edge[above, bend left] node{$a$} (y)
	(y) edge[below, bend left] node{$b$} (x)
	;
	\end{tikzpicture}}	
			\caption{The minimal DFA}
	\label{m(l)}
\end{subfigure}
\begin{subfigure}[b]{0.39 \textwidth}
		\centering	
		\tiny
								\adjustbox{valign=m}{
			\begin{tikzpicture}[node distance=6em]
				\node[state] (0) {$\lbrack \emptyset \rbrack$};
				\node[state, right of=0, initial, initial text=] (x) {$\lbrack \lbrace x \rbrace \rbrack$};
					\node[state, right of=x, accepting] (y) {$\lbrack \lbrace y \rbrace \rbrack$};	
			    \path[->]
			(0) edge[loop above] node{$a,b$} (0)
			(x) edge[loop above] node{$b$} (x)
			(x) edge[above, bend left] node{$a$} (y)
			(y) edge[below, bend left] node{$b$} (x)
			(y) edge[loop right] node{$a$} (y)
			;
			\end{tikzpicture}
			}
			\adjustbox{valign=m}{
			\resizebox{0.65 \columnwidth}{!}{%		
				\begin{tabular}{ c|c|c|c } 
 $\vee$ & $\lbrack \lbrace x \rbrace \rbrack$ & $ \lbrack \lbrace y \rbrace \rbrack$ &  $\lbrack \emptyset \rbrack$ \\
 \hline 
$\lbrack \lbrace x \rbrace \rbrack$ & $\lbrack \lbrace x \rbrace \rbrack$ & $\lbrack \lbrace y \rbrace \rbrack$ & $\lbrack \lbrace x \rbrace \rbrack$ \\ 
 \hline
 $\lbrack \lbrace y \rbrace \rbrack$ & $\lbrack \lbrace y \rbrace \rbrack$ & $\lbrack \lbrace y \rbrace \rbrack$ & $\lbrack \lbrace y \rbrace \rbrack$ \\
 \hline
 $\lbrack \emptyset \rbrack$ & $\lbrack \lbrace x \rbrace \rbrack$ & $\lbrack \lbrace y \rbrace \rbrack$ & $\lbrack \emptyset \rbrack$
\end{tabular}}}
					\caption{The minimal CSL-structured DFA}
		\label{overlineml}
		\end{subfigure}
	\begin{subfigure}[b]{.26 \textwidth}
		\tiny
		\centering
							\adjustbox{valign=m}{
		\begin{tikzpicture}[node distance=6em]
			\node[state, initial, initial text=] (x) {$\lbrack \lbrace x \rbrace \rbrack$};
				\node[state, right of=x, accepting] (y) {$\lbrack \lbrace y \rbrace \rbrack$};	
		    \path[->]
		(x) edge[loop above] node{$a,b$} (x)
		(x) edge[above, bend left] node{$a$} (y)
		(y) edge[below, bend left] node{$a,b$} (x)
		(y) edge[loop right] node{$a$} (y)
		;
		\end{tikzpicture}
		}
		\caption{The canonical RFSA}
	\label{jiromaton}
	\end{subfigure}	
	\caption{Three automata accepting the language $(a+b)^*a \subseteq \lbrace a, b \rbrace ^*$.}
	\label{crfsadiag1}
\end{figure}

\section{Preliminaries}

We assume basic knowledge of category theory (including functors, natural transformations, adjunctions), for an overview see e.g.  \cite{awodey2010category}. 

We briefly recall the definitions of coalgebras, monads, and Eilenberg-Moore algebras. A \emph{coalgebra} for an endofunctor $F$ on a category $\C$ is a tuple $(X, k)$ consisting of an object $X$ in $\C$ and a morphism $k\colon X \rightarrow FX$. A homomorphism $f: (X,k_X) \rightarrow (Y, k_Y)$ between coalgebras for $F$ is a morphism $f: X \rightarrow Y$ in $\C$ satisfying $k_Y \circ f = Ff \circ k_X$. The category of coalgebras for $F$ and homomorphisms is denoted by $\textnormal{Coalg}(F)$. 

A \emph{monad} on a category $\mathscr{C}$ is a tuple $(T, \eta, \mu)$ consisting of an endofunctor $T: \mathscr{C} \rightarrow \mathscr{C}$ and natural transformations
	$
	\eta: \textnormal{id}_{\mathscr{C}} \Rightarrow T$ and $\mu: T^2 \Rightarrow T$
	satisfying $\mu \circ T\mu = \mu \circ \mu_T$ and $\mu \circ \eta_T = \textnormal{id}_T = \mu \circ T\eta$. A morphism $(F, \alpha): (\mathscr{C}, S) \rightarrow (\mathscr{D}, T)$  between a monad $S$ on a category $\mathscr{C}$ and a monad $T$ on a category $\mathscr{D}$ consists of a functor $F: \mathscr{C} \rightarrow \mathscr{D}$ and a natural transformation $\alpha: TF \Rightarrow FS$ satisfying $\alpha \circ \eta^T F = F\eta^S$ and $F\mu^S \circ \alpha S \circ T \alpha = \alpha \circ \mu^T F$ \cite{STREET1972149}. 
		The composition of two monad morphisms $(F, \alpha): (\mathscr{C}, S) \rightarrow (\mathscr{D}, T)$ and $(G, \beta): (\mathscr{D}, T) \rightarrow (\mathscr{E}, U)$ is the monad morphism $(GF, G\alpha \circ \beta F): (\mathscr{C}, S) \rightarrow (\mathscr{E}, U)$ \cite{STREET1972149}. Two well-known monads on the category of sets and functions are the \emph{powerset monad} $\mathcal{P}$ and the \emph{free $\mathbb{K}$-vector space monad} $\mathcal{V}_{\mathbb{K}}$ \cite{zetzsche2021}.

	An \emph{Eilenberg-Moore algebra} over a monad $T$ on $\C$ is a tuple $(X, h)$ consisting of an object $X$ in $\C$ and a morphism $h: TX \rightarrow X$ satisfying $h \circ \mu_X = h \circ Th$ and $h \circ \eta_X = \textnormal{id}_X$. A homomorphism $f: (X,h_X) \rightarrow (Y, h_Y)$ between Eilenberg-Moore algebras over $T$ is a morphism $f: X \rightarrow Y$ in $\C$ satisfying $h_Y \circ Tf = f \circ h_X$. The category of Eilenberg-Moore algebras over $T$ is denoted by $\EM$. One can show that the category of algebras over $\mathcal{P}$ is isomorphic to the category of complete join-semi lattices, and the category of algebras over $\mathcal{V}_{\mathbb{K}}$ is isomorphic to the category of $\mathbb{K}$-vector spaces.
	
We now introduce other notions that are necessary to follow our technical development: distributive laws, bialgebras, and generators and bases for algebras over a monad.

Distributive laws have originally occurred as a way to compose monads \cite{beck1969distributive}, but now also exist in a wide range of other forms \cite{Street2009}. For our case it is sufficient to consider distributive laws between a monad and an endofunctor, sometimes called \emph{Eilenberg-Moore laws} \cite{jacobs2012trace}.

\begin{definition}[Distributive Law]
\label{distributivelawdef}
	A \emph{distributive law} between a monad $T$  and an endofunctor $F$ on $\C$ is a natural transformation $\lambda: TF \Rightarrow FT$
satisfying $F\eta_X = \lambda_X \circ \eta_{FX}$ and $\lambda_X \circ \mu_{FX} = F\mu_X \circ \lambda_{TX} \circ T\lambda_X$.
\end{definition}

Given a distributive law, one can model the determinisation of a system with dynamics in $F$ and side-effects in $T$ by lifting a $FT$-coalgebra $(X, k)$ to the $F$-coalgebra 
	$(TX, k^{\sharp})$, where $k^{\sharp} := (F \mu_X \circ \lambda_{TX}) \circ Tk$. As one verifies, $k^{\sharp}$ is a $T$-algebra homomorphism of type $(TX, \mu_X) \rightarrow (FTX, F\mu_X \circ \lambda_{TX})$. There exists a distributive law for which the lifting $k^{\sharp}$ is the DFA in CSL obtained from an NFA $k$ via the classical powerset construction \cite{silva2010generalizing}.   
	
		The example illustrates the concept of a bialgebra: the algebraic part $(TX, \mu_X)$ and the coalgebraic part $(TX, k^{\sharp})$ of a lifted automaton are compatible along the distributive law $\lambda$.
    \begin{definition}[Bialgebra]
    \label{bialgebradef}
		A $\lambda$\emph{-bialgebra} is a tuple $(X, h, k)$ consisting of a $T$-algebra $(X,h)$ and an $F$-coalgebra $(X, k)$ 
satisfying $Fh \circ \lambda_X \circ Tk = k \circ h$.
	\end{definition}
	
		 A homomorphism between $\lambda$-bialgebras is a morphism between the underlying objects that is simultaneously a $T$-algebra homomorphism and an $F$-coalgebra homomorphism.
	The category of $\lambda$-bialgebras and homomorphisms is denoted by $\textnormal{Bialg}(\lambda)$. 

The generalised determinisation can be rephrased as a functor $\textnormal{exp}_T$ that \emph{expands} a $F$-coalgebra with side-effects in $T$ into a $\lambda$-bialgebra. We will also refer to the functor $\textnormal{free}_T$ that arises from $\textnormal{exp}_T$ by pre-composition with the canonical embedding of $F$-coalgebras into $FT$-coalgebras, therefore assigning to a $F$-coalgebra the $\lambda$-bialgebra it \emph{freely} generates.

 \begin{lemma}[\cite{jacobs2012trace}]
	\label{expfunctor}
	\begin{itemize}
		\item Defining $\textnormal{exp}_T(X, k) := (TX, \mu_X, F \mu_X \circ \lambda_{TX} \circ Tk)$ and $\textnormal{exp}_T(f) := Tf$ yields a functor $\textnormal{exp}_T: \textnormal{Coalg}(FT) \rightarrow \textnormal{Bialg}(\lambda)$. 
		\item Defining $\textnormal{free}_T(X, k) := (TX, \mu_X, \lambda_X \circ Tk)$ and $\textnormal{free}_T(f) := Tf$ yields a functor $\textnormal{free}_T: \textnormal{Coalg}(F) \rightarrow \textnormal{Bialg}(\lambda)$ satisfying $\textnormal{free}_T(X, k) = \textnormal{exp}_T(X, F\eta_X \circ k)$.
	\end{itemize}
	\end{lemma}
	
	The last ingredient is a generalisation of generators for structures such as vector spaces.
	
\begin{definition}[Generator and Basis \cite{zetzsche2021}]
\label{generatordefinition}
	A \emph{generator} for a $T$-algebra $( X, h )$ is a tuple $( Y, i, d )$  consisting of an object $Y$, a morphism $i \colon Y \rightarrow X$, and a morphism $d \colon X \rightarrow TY$ such that $(h \circ Ti) \circ d = \textnormal{id}_X$.
	A generator is called a \emph{basis} if it additionally satisfies $d \circ (h \circ Ti) = \textnormal{id}_{TY}$.
\end{definition}

A generator for a $T$-algebra is called a \textit{scoop} by Arbib and Manes~\cite{arbib1975fuzzy}. 
Intuitively, a set $Y$ embedded into an algebraic structure $X$ via $i$ is a generator for the latter if every element $x \in X$ admits a decomposition into a \emph{formal} combination $d(x) \in TY$ of elements of $Y$ that evaluates to $x$ via the interpretation $h \circ Ti$.
If the decomposition is moreover \emph{unique}, that is, $h \circ Ti$ is not only a \emph{surjection} with right-inverse $d$, but a \emph{bijection} with two-sided inverse $d$, then a generator is called a basis.
Every $T$-algebra $(X,h)$ is generated by $( X, \textnormal{id}_X, \eta_X )$ and admits a basis iff it is isomorphic to a free algebra.

%Intuitively, a subset $Y$ of an algebraic structure $X$ is a generator for the latter, if every $x \in X$ decomposes into a formal combination $d(x) \in TY$  of elements in $Y$ that evaluates to $x$.
%If the decomposition is moreover \emph{unique}, that is, $h \circ Ti$ is not only a \emph{surjection} with right-inverse $d$, but a \emph{bijection} with two-sided inverse $d$, then a generator is called a basis.

 \begin{example}
\label{setbasedexamples}
\begin{itemize}
	\item A tuple $( Y, i, d )$ is a generator for a $\mathcal{P}$-algebra $L = ( X,h ) \simeq ( X, \vee^h )$ iff $x = \vee^h_{y \in d(x)} i(y) $ for all $x \in X$, where we write $\vee^h$ for the complete join-semilattice structure induced by $h$. In the case that $Y \subseteq X$ is a subset, one typically defines $i(y) = y$ for all $y \in Y$.
            If $L$ satisfies the descending chain condition, which is in particular the case if $X$ is finite, then defining $i(y) = y$ and $d(x) = \lbrace y \in J(L) \mid y \leq x \rbrace$ turns the set of join-irreducibles $J(L)$ into a size-minimal generator $( J(L), i, d )$ for $L$ \cite{zetzsche2021}.
     \item A tuple $( Y, i, d )$ is a generator for a $\mathcal{V}_{\mathbb{K}}$-algebra $V = ( X,h ) \simeq ( X, +^h, \cdot^h )$ iff $x = \sum^h_{y \in Y} d(x)(y) \cdot^h i(y)$ for all $x \in X$, where we write $+^h$ and $\cdot^h$ for the $\mathbb{K}$-vector space structure induced by $h$.
			As every vector space can be equipped with a basis, every $\mathcal{V}_{\mathbb{K}}$-algebra $V$ admits a basis. One can show that a basis is a size-minimal generator \cite{zetzsche2021}.
\end{itemize}			
\end{example}

A central result in \cite{zetzsche2021} shows that it is enough to find generators for the underlying algebra of a bialgebra to derive an equivalent free bialgebra. This is because the algebraic and coalgebraic components are tightly intertwined via a distributive law.

\begin{proposition}[{\cite{zetzsche2021}}]
\label{forgenerator-isharp-is-bialgebra-hom}
	Let $( X, h, k)$ be a $\lambda$-bialgebra and let $( Y, i, d )$ be a generator for the $T$-algebra $( X,h )$.
	Then $h \circ Ti \colon \textnormal{exp}_T( Y, Fd \circ k \circ i)  \rightarrow ( X, h, k )$ is a $\lambda$-bialgebra homomorphism.
\end{proposition}

\section{Step 1: Closure}

\label{closure_sec}

In this section, we further explore the categorical construction of minimal canonical acceptors given in \cite{zetzsche2021}. In particular, we show that deriving a minimal bialgebra from a minimal coalgebra by closing the latter with additional algebraic structure has a direct analogue in universal algebra: taking the closure of a subset of an algebra.

\subsection{Factorisation Systems and Subobjects}

In the category of sets and functions, every morphism can be factored into a surjection onto its image followed by an injection into the codomain of the morphism. In this section we recall a convenient abstraction of this phenomenon for arbitrary categories. The ideas are well established \cite{bousfield1977constructions,riehl2008factorization,maclane1950duality}. We choose to adapt the formalism of \cite{adamek2009abstract}.

\begin{definition}[Factorisation System]
\label{factorisation_system_def}
Let $\mathscr{E}$ and $\mathscr{M}$ be classes of morphisms in a category $\mathscr{C}$. We call the tuple $(\mathscr{E}, \mathscr{M})$ a \emph{factorisation system} for $\mathscr{C}$ if the following three conditions hold:
\begin{enumerate}[leftmargin=1cm]
	\item[(F1)] Each of $\mathscr{E}$ and $\mathscr{M}$ is closed under composition with isomorphisms.
	\item[(F2)] Each morphism $f$ in $\mathscr{C}$ can be factored as $f = m \circ e$, with $e \in \mathscr{E}$ and $m \in \mathscr{M}$.
	\item[(F3)] Whenever $g \circ e = m \circ f$ with $e \in \mathscr{E}$ and $m \in \mathscr{M}$, there exists a unique diagonal $d$, such that $f = d \circ e$ and $g = m \circ d$. 
%	
%	
%	For each commutative square with $e \in \mathscr{E}$ and $m \in \mathscr{M}$ as on the left, there exists a unique diagonal $d$ such that the diagram on the right commutes:	 \begin{equation*}
%			\begin{tikzcd}
%				\cdot \arrow{r}{e} \arrow{d}[left]{f} &\cdot \arrow{d}{g} \\
%				\cdot \arrow{r}[below]{m} &\cdot
%			\end{tikzcd}
%			\qquad
%			\begin{tikzcd}
%							\cdot \arrow{r}{e} \arrow{d}[left]{f} &\cdot \arrow{d}{g} \arrow[dashed]{dl}{d} \\
%				\cdot \arrow{r}[below]{m} &\cdot
%			\end{tikzcd}
%		\end{equation*}
\end{enumerate}
\end{definition}

We use double headed ($\twoheadrightarrow$) and hooked ($\hookrightarrow$) arrows to indicate that a morphism is in $\mathscr{E}$ or $\mathscr{M}$, respectively. If $f$ factors into $e$ and $m$, we call the codomain of $e$, or equivalently, the domain of $m$, the \emph{image} of $f$ and denote it by $\textnormal{im}(f)$.

One can show that each of $\mathscr{E}$ and $\mathscr{M}$ contains all isomorphisms and is closed under composition \cite[Prop. 14.6]{adamek2009abstract}.
From the uniqueness of the diagonal one can deduce that factorisations are unique up to unique isomorphism \cite[Prop. 14.4]{adamek2009abstract}. 
It further follows that $\mathscr{E}$ has the \emph{right cancellation property}, that is $g \circ f \in \mathscr{E}$ and $f \in \mathscr{E}$ implies $g \in \mathscr{E}$. Dually, $\mathscr{M}$ has the \emph{left cancellation property}, that is, $g \circ f \in \mathscr{M}$ and $g \in \mathscr{M}$ implies $f \in \mathscr{M}$ \cite[Prop. 14.9]{adamek2009abstract}. 

As intended, in the category of sets and functions, surjective and injective functions, or equivalently, epi- and monomorphisms, constitute a factorisation system \cite[Ex. 14.2]{adamek2009abstract}. More involved examples can be constructed for e.g. the category of topological spaces or the category of categories \cite[Ex. 14.2]{adamek2009abstract}. We are particularly interested in factorisation systems for the categories of algebras over a monad and coalgebras over an endofunctor.

The naive categorification of a subset $Y \subseteq X$ is a monomorphism $Y \rightarrow X$. Since in the category of sets epi- and monomorphism constitute a factorisation system, we may generalise subsets to arbitrary categories $\mathscr{C}$ with a factorisation system $(\mathscr{E}, \mathscr{M})$ in the following way:  

\begin{definition}[Subobjects]
\label{subobject_def}
	A \emph{subobject} of an object $X \in \mathscr{C}$ is a morphism $m_Y: Y \hookrightarrow X \in \mathscr{M}$. A morphism $f: m_{Y_1} \rightarrow m_{Y_2}$ between subobjects of $X$ consists of a morphism $f: Y_1 \rightarrow Y_2$ such that $m_{Y_2} \circ f = m_{Y_1}$. 
\end{definition}

The category of (isomorphism classes of) subobjects of $X$ is denoted by $\textnormal{Sub}(X)$.

As $\mathscr{M}$ has the left cancellation property, every morphism between subobjects in fact lies in $\mathscr{M}$. We work with isomorphism classes of subobjects since factorisations of morphisms are only defined up to unique isomorphism. For epi-mono factorizations, there is at most one morphism between any two subobjects, that is, $\textnormal{Sub}(X)$ is simply a partially ordered set.

 \label{factorisationsystemalgebra}

\begin{figure*}
\centering
		\begin{tikzcd}[row sep = 0.95em]
		TX \arrow{d}[left]{h_X} \arrow{r}{Tf} & TY \arrow{d}{h_Y} \\
		X \arrow{r}[below]{f} & Y
	\end{tikzcd}
	\qquad
	\begin{tikzcd}[row sep = 0.95em]
		X \arrow[twoheadrightarrow]{r}{e}  \arrow{dr}[below]{f} & \textnormal{im}(f) \arrow[hookrightarrow]{d}{m} \\
		  & Y 
	\end{tikzcd}
	\qquad
			\begin{tikzcd}[row sep = 0.95em]
		TX \arrow{d}[left]{e \circ h_X} \arrow[twoheadrightarrow]{r}{Te} & T\textnormal{im}(f) \arrow{d}{h_Y \circ Tm} \arrow[dashed]{dl}{h_{\textnormal{im}(f)}} \\
		\textnormal{im}(f) \arrow[hookrightarrow]{r}[below]{m} & Y
	\end{tikzcd}
\caption{Factorising a $T$-algebra homomorphism via the factorisation system of a base category.}
\label{monadpreserveepilift}
\end{figure*}

\subsection{Factorising Algebra Homomorphisms}

\label{factorisingalgebrahom}

In this section, we recall that if one is given a category $\mathscr{C}$ with a factorisation system $(\mathscr{E}, \mathscr{M})$ and a monad $T$ on $\mathscr{C}$ that preserves $\mathscr{E}$ (that is, satisfies $T(e) \in \mathscr{E}$ for all $e \in \mathscr{E}$), it is possible to lift the factorisation system of the base category $\mathscr{C}$ to a factorisation system on the category of Eilenberg-Moore algebras $\EM$. 

The result appears in e.g \cite{wissmann2022minimality} and may be extended to algebras over an endofunctor. It can also be stated in its dual version: if an endofunctor on $\mathscr{C}$ preserves $\mathscr{M}$, it is possible to lift the factorisation system of $\mathscr{C}$ to the category of coalgebras \cite{kurzlogics, wissmann2022minimality}. 

The induced factorisation system for $\EM$ consists of those algebra homomorphisms, whose underlying morphism lies in $\mathscr{E}$ or $\mathscr{M}$, respectively. Clearly in such a system condition (F1) holds. The next result shows that it also satisfies (F3).

\begin{restatable}[{\cite[Lem. 3.6]{wissmann2022minimality}}]{lemma}{diagonalalgebras}
\label{diagonalalgebras}
Whenever $g \circ e = m \circ f$ for $T$-algebra homomorphisms $f,g,e,m$, with $e \in \mathscr{E}$ and $m \in \mathscr{M}$, there exists a unique diagonal $T$-algebra homomorphism $d$, such that $f = d \circ e$ and $g = m \circ d$. 

%For each commutative square of homomorphisms between algebras for the endofunctor $T$ as on the left below
%\begin{equation*}
%		\begin{tikzcd}[ampersand replacement=\&]
%	(A, h_A) \arrow[twoheadrightarrow]{r}{e} \arrow{d}[left]{f} \& (B, h_B) \arrow{d}{g}  \\
%	(C, h_C) \arrow[hookrightarrow]{r}[below]{m} \&(D, h_D)
%\end{tikzcd}
%\qquad
%	\begin{tikzcd}[ampersand replacement=\&]
%	(A, h_A) \arrow[twoheadrightarrow]{r}{e} \arrow{d}[left]{f} \& (B, h_B) \arrow{d}{g} \arrow[dashed]{dl}{d} \\
%	(C, h_C) \arrow[hookrightarrow]{r}[below]{m} \&(D, h_D)
%\end{tikzcd}
%\end{equation*}
%there exists a unique diagonal $d: (B, h_B) \rightarrow (C, h_C)$ such that the diagram on the right above commutes.
\end{restatable}	

Let us now show that the proposed factorisation system satisfies (F2). Assume we are given a homomorphism $f$ as on the left of \Cref{monadpreserveepilift}. Using the factorisation system of the base category $\mathscr{C}$, we can factorise it, as ordinary morphism, into $e \in \mathscr{E}$ and $m \in \mathscr{M}$. In consequence the outer square of the diagram on the right of \Cref{monadpreserveepilift} commutes. Since by assumption the morphism $Te$ is again in $\mathscr{E}$, we thus find a unique diagonal $h_{\textnormal{im}(f)}$ in $\mathscr{C}$ that makes the triangles on the right of \Cref{monadpreserveepilift}  commute. The result below shows that $h_{\textnormal{im}(f)}$ equips $\textnormal{im}(f)$ with the structure of a $T$-algebra.

\begin{restatable}[{\cite[Prop. 3.7]{wissmann2022minimality}}]{lemma}{imisalgebra}
\label{imisalgebra}
	$(\textnormal{im}(f), h_{\textnormal{im}(f)})$ is an Eilenberg-Moore $T$-algebra.
\end{restatable}

We thus obtain a factorisation of $f: (X, h_X) \rightarrow (Y,h_ Y)$ into Eilenberg-Moore $T$-algebra homomorphisms $e: (X,h_X) \twoheadrightarrow (\textnormal{im}(f), h_{\textnormal{im}(f)})$ and $m: (\textnormal{im}(f), h_{\textnormal{im}(f)}) \hookrightarrow (Y, h_Y)$.

\subsection{The Subobject Closure Functor}

\label{closureasfunctorsec}

While subobjects in the category of sets generalise subsets, subobjects in the category of algebras generalise subalgebras. By taking the algebraic closure of a subset of an algebra one can thus transition from one category of subobjects to the other. 

In this section, we generalise this phenomenon from the category of sets to more general categories. As before, we assume a base category $\mathscr{C}$ with a factorisation system $(\mathscr{E}, \mathscr{M})$ and a monad $T$ on $\mathscr{C}$ that preserves $\mathscr{E}$. Our aim is to construct, for any $T$-algebra $\mathbb{X}$ with carrier $X$, a functor from the subobjects $\textnormal{Sub}(X)$ in $\mathscr{C}$ to the subobjects $\textnormal{Sub}(\mathbb{X})$ in $\EM$ that assigns to a subobject of $X$ its \emph{closure}, that is, the least $T$-subalgebra of $\mathbb{X}$ containing it.

 \begin{figure}[t]
\centering
\begin{subfigure}[b]{0.45 \textwidth}
\centering
	\begin{tikzcd}[row sep = 0.75em]
 			\mathscr{C}/X \arrow{r}{} & \EM/\mathbb{X} \arrow{d}{} \\
 			\textnormal{Sub}(X) \arrow{u}{} \arrow{r}[above]{\overline{(\cdot)}^{\mathbb{X}}} & \textnormal{Sub}(\mathbb{X})
 		\end{tikzcd}	
\caption{Decomposition}
\label{closure_as_composition}
\end{subfigure}
\begin{subfigure}[b]{0.45 \textwidth}
\centering
  		\begin{tikzcd}[row sep = 0.75em]
 			\mathscr{C}/X \arrow{r}{} \arrow{d}[left]{}  & \EM/\mathbb{X} \arrow{d}{} \\
 			\textnormal{Sub}(X) \arrow{r}[above]{\overline{(\cdot)}^{\mathbb{X}}} & \textnormal{Sub}(\mathbb{X})
 		\end{tikzcd}
\caption{Commutativity}
\label{closure_commuting}
 \end{subfigure}	
 \caption{A high-level perspective on the subobject closure functor defined in \Cref{functorprop}.}
\end{figure}

  Recall the free Eilenberg-Moore algebra adjunction. For any object $Y$ in $\mathscr{C}$ and $T$-algebra $\mathbb{X} = (X,h)$, it maps a morphism $\varphi: Y \rightarrow X$ to the $T$-algebra homomorphism $\varphi^{\sharp} := h \circ T\varphi : (TY, \mu_Y) \rightarrow \mathbb{X}$. In \Cref{factorisationsystemalgebra} we have seen that the factorisation system of $\mathscr{C}$ naturally lifts to a factorisation system on the category of $T$-algebras. In particular, we know that up to isomorphism the homomorphism $\varphi^{\sharp}$ admits a factorisation into algebra homomorphisms of the form $\varphi^{\sharp} = m_{\textnormal{im}(\varphi^{\sharp})} \circ e_{\textnormal{im}(\varphi^{\sharp})}$.
If the morphism $\varphi$ is given by a subobject $m_Y$, let $\overline{Y} := (\textnormal{im}(m_Y^{\sharp}), h_{\textnormal{im}(m_Y^{\sharp})})$, then above construction yields a second subobject $m_{\overline{Y}}$:
 \[ m_Y: Y \rightarrow X \in \mathscr{M} \qquad \qquad m_{\overline{Y}}: \overline{Y} \rightarrow \mathbb{X} \in \mathscr{M}. \] 

Since for any morphism $f: m_{Y_1} \rightarrow m_{Y_2}$ between subobjects of $X$ one has $m_{\overline{Y_1}} \circ e_{\overline{Y_1}} = m_{\overline{Y_2}} \circ (e_{\overline{Y_2}} \circ Tf)$, there exists a unique homomorphism $\overline{f}: m_{\overline{Y_1}} \rightarrow m_{\overline{Y_2}}$ satisfying $\overline{f} \circ e_{\overline{Y_1}} = e_{\overline{Y_2}} \circ Tf$.
 
 The following result shows that above constructions are compositional.

\begin{restatable}{proposition}{functorprop}
\label{functorprop}
Assigning $m_Y \mapsto m_{\overline{Y}}$ and $f \mapsto \overline{f}$ yields a functor $\overline{(\cdot)}^{\mathbb{X}}: \textnormal{Sub}(X) \rightarrow \textnormal{Sub}(\mathbb{X})$.
\end{restatable}

%   Let us instantiate above result for the free vector space monad on the category of sets equipped with its canonical surjective-injective factorisation system.
% 
% \begin{example}
% \label{freevectorspaceexamplefunctor}
% Given an injective embedding $m_Y$ of some set $Y$ into a vector space $\mathbb{V}$, one easily verifies that the lifting $(m_Y)^{\sharp}$ maps a formal linear combination $\sum_i \lambda_i \cdot y_i$ to the vector $\sum_i \lambda_i \cdot m_Y(y_i)$ in $V$. The image $\overline{Y}$ is thus given by the vector space that consists of equivalence classes of formal linear combinations, and the injection $m_{\overline{Y}}$ interprets representatives as demonstrated. In particular, if $Y$ is a subset of $\mathbb{V}$, that is, $m_Y(y) = y$ for all $y \in Y$, the closure can be recognised as the so-called sub vector space of $\mathbb{V}$ generated by $Y$.
% \end{example}
   
Mapping an algebra homomorphism with codomain $\mathbb{X}$ to the $\mathscr{M}$-part of its factorisation extends to a functor from the slice category
over $\mathbb{X}$ (in which we here and in the following identify isomorphic objects) to the category of subobjects of $\mathbb{X}$.   
Similarly, one observes that the free Eilenberg-Moore algebra adjunction gives rise to a functor from the slice category over $X$ to the slice category over $\mathbb{X}$.  Finally, it is clear that there exists a functor from the category of subobjects of $X$ to the slice category over $X$. The functor defined in \Cref{functorprop} can thus be recognised as the composition in \Cref{closure_as_composition}.

 \subsection{The Subobject Closure Monad}
 
\label{closureasmonadsec} 
 
 In this section, we show that the functor in \Cref{functorprop} induces a monad on the category of subobjects $\textnormal{Sub}(X)$. 
 As before, we assume a base category $\mathscr{C}$ with a factorisation system $(\mathscr{E}, \mathscr{M})$ and a monad $T = (T, \eta, \mu)$ on $\mathscr{C}$ that preserves $\mathscr{E}$. 

 We begin by establishing the following two technical identities, which assume a $T$-algebra $\mathbb{X} = (X,h)$ and a subobject $m_Y: Y \rightarrow X \in \mathscr{M}$.
 
\begin{restatable}{lemma}{twoequalities}
\label{twoequalities}
$m_{\overline{Y}} \circ e_{\overline{Y}} \circ \eta_Y = m_Y$
	and $m_{\overline{Y}} \circ e_{\overline{Y}} \circ \mu_Y = m_{\overline{\overline{Y}}} \circ e_{\overline{\overline{Y}}} \circ Te_{\overline{Y}}$.
\end{restatable}

  In consequence, we can define candidates for the monad unit $\eta^{\mathbb{X}}$ and the monad multiplication $\mu^{\mathbb{X}}$, respectively, as the unique diagonals in \Cref{unitmultdef}.
 By construction both morphisms are homomorphisms of subobjects: $\eta^{\mathbb{X}}_{m_Y}: m_Y \rightarrow m_{\overline{Y}}$ and $\mu^{\mathbb{X}}_{m_Y}: m_{\overline{\overline{Y}}} \rightarrow m_{\overline{Y}}$. 
  The remaining proofs of naturality and the monad laws are covered below. By a slight abuse of notation, we write $\overline{(\cdot)}^{\mathbb{X}}$ for the endofunctor on $\textnormal{Sub}(X)$ that arises by post-composition of the functor in \Cref{functorprop} with the canonical forgetful functor from $\textnormal{Sub}(\mathbb{X})$ to $\textnormal{Sub}(X)$. 
  
  \begin{restatable}{theorem}{inducedmonad}
   \label{inducedmonad}
   	$(\overline{(\cdot)}^{\mathbb{X}}, \eta^{\mathbb{X}}, \mu^{\mathbb{X}})$ is a monad on $\textnormal{\textnormal{Sub}}(X)$.	
  \end{restatable}

%
% 
%As before, we exemplify \Cref{inducedmonad} for the  free vector space monad on the category of sets and functions with its canonical factorisation system.
%
%\begin{example}
%We have previously seen that the the closure of a subobject $m_Y$ of a vector space $\mathbb{V}$ consists of formal linear combinations that are considered equivalent, if their interpretation in $\mathbb{V}$ via $m_Y$ coincides. By construction (cf. \eqref{unitmultdef}), the monad unit $\eta^{\mathbb{V}}_{m_Y}$ maps an element $y \in Y$ to the equivalence class $\lbrack 1 \cdot y \rbrack$ in $\overline{Y}$. One further verifies that the multiplication $\mu^{\mathbb{V}}_{m_Y}$ assigns to the element $\lbrack \sum_{\lbrack \varphi \rbrack} \lambda_{\lbrack \varphi \rbrack} \cdot \lbrack \varphi \rbrack \rbrack$ in $\overline{\overline{Y}}$  the element $\lbrack \sum_x (\sum_{\lbrack \varphi \rbrack} \lambda_{\lbrack \varphi \rbrack} \cdot \varphi(y)) \cdot y \rbrack$ in $\overline{Y}$. If $Y$ is a subset of $\mathbb{V}$, that is, $m_Y(y) = y$ for all $y \in Y$, the multiplication thus flattens vectors of vectors in the usual way.
%\end{example}

We will now show that the mapping of an algebra $\mathbb{X}$ to the monad $\overline{(\cdot)}^{\mathbb{X}}$ in \Cref{inducedmonad} extends to algebra homomorphisms. To this end, for any algebra homomorphism $f: \mathbb{A} \rightarrow \mathbb{B}$ in $\mathscr{M}$, let 
$f_{*}: \textnormal{Sub}(A) \rightarrow \textnormal{Sub}(B)$ be the induced functor defined by $f_{*}(m_X) = f \circ m_X$ and $f_*(g) = g$. The result below shows that $f_*$ can be extended to a morphism between monads. 

  \begin{restatable}{lemma}{morphismmonadone}
 \label{morphismmonadone}
	For any $f: \mathbb{A} \rightarrow \mathbb{B} \in \mathscr{M}$, there exists a monad morphism $(f_{*}, \alpha): (\textnormal{Sub}(A), \overline{(\cdot)}^{\mathbb{A}}) \rightarrow (\textnormal{Sub}(B), \overline{(\cdot)}^{\mathbb{B}})$. 	
  \end{restatable}

 The next statement establishes that the canonical forgetful functor $U: \textnormal{Sub}(X) \rightarrow \mathscr{C}$ defined by $U(m_Y) = Y$ and $U(f) = f$ extends to a morphism between monads.
   \begin{restatable}{lemma}{morphismmonadtwo}
   	  \label{morphismmonadtwo}
 	There exists a monad morphism $(U, \alpha): (\textnormal{Sub}(X), \overline{(\cdot)}^{\mathbb{X}}) \rightarrow (\mathscr{C}, T)$.
   \end{restatable}

We conclude with the observation that the monad morphism defined in \Cref{morphismmonadone} commutes with the monad morphisms defined in \Cref{morphismmonadtwo}.

  \begin{restatable}{lemma}{monadmorpismcommuting}
  \label{monadmorpismcommuting}
  	\Cref{monadmorphismdiagram} commutes for any algebra homomorphism $f: \mathbb{A} \rightarrow \mathbb{B} \in \mathscr{M}$. 
  \end{restatable}

   \begin{figure*}[t]
   \centering
\begin{subfigure}{0.48 \textwidth}
 		\begin{tikzcd}[row sep =0.9em]
 			Y \arrow[twoheadrightarrow]{r}{1} \arrow{d}[left]{e_{\overline{Y}} \circ \eta_Y} & Y \arrow[dashed]{dl}{\eta^{\mathbb{X}}_{m_Y}} \arrow{d}{m_Y} \\
 			\overline{Y} \arrow[hookrightarrow]{r}[below]{m_{\overline{Y}}} & X
 		\end{tikzcd} 
\begin{tikzcd}[row sep =0.9em]
 			T^2Y \arrow[twoheadrightarrow]{r}{e_{\overline{\overline{Y}}} \circ Te_{\overline{Y}}} \arrow{d}[left]{e_{\overline{Y}} \circ \mu_Y} & \overline{\overline{Y}} \arrow[dashed]{dl}{\mu^{\mathbb{X}}_{m_Y}} \arrow{d}{m_{\overline{\overline{Y}}}} \\
 			\overline{Y} \arrow[hookrightarrow]{r}[below]{m_{\overline{Y}}} & X
 		\end{tikzcd}
\caption{Induced unit $\eta^{\mathbb{X}}$ and multiplication $\mu^{\mathbb{X}}$ of the monad in \Cref{inducedmonad}.}
  \label{unitmultdef}
    \end{subfigure}
    \quad
\begin{subfigure}{0.4 \textwidth}
\centering
 \begin{tikzcd}[ampersand replacement=\&, row sep =0.9em, column sep = 0.4em]
 	(\textnormal{Sub}(A), \overline{(\cdot)}^{\mathbb{A}}) \arrow{dr}[left]{(U_{\mathbb{A}}, \alpha_{\mathbb{A}})} \arrow{rr}{(f_*, \alpha_f)} \& \& (\textnormal{Sub}(B), \overline{(\cdot)}^{\mathbb{B}}) \arrow{dl}{(U_{\mathbb{B}}, \alpha_{\mathbb{B}})} \\
 	\& (\mathscr{C}, T) \&
 \end{tikzcd}
 \caption{Commutativity of the monad morphisms in \Cref{morphismmonadone} and \Cref{morphismmonadtwo}.}
 \label{monadmorphismdiagram}
\end{subfigure}
\caption{Structure and properties of the monad in \Cref{inducedmonad}.}
  \end{figure*}

 \subsection{Closing an Image}
 
 \label{closinganimagesec}
 
 In this section we investigate the closure of a particular class of subobjects: the ones that arise by taking the image of a morphism. We then show that deriving a minimal bialgebra from a minimal coalgebra by equipping the latter with additional algebraic structure can be realized as the closure of a subobject in this class.
 
  As before, we assume a category $\mathscr{C}$ with a factorisation system $(\mathscr{E}, \mathscr{M})$ and a monad $T$ on $\mathscr{C}$ that preserves $\mathscr{E}$. Suppose that $\mathbb{X} = (X, h_X)$ is a $T$-algebra and $f: Y \rightarrow X$ a morphism in $\mathscr{C}$.
On the one hand, there exists a factorisation of $f$ in $\mathscr{C}$:
 \[
 f = Y \overset{e_{\textnormal{im}(f)}}{\twoheadrightarrow} \textnormal{im}(f) \overset{m_{\textnormal{im}(f)}}{\hookrightarrow} X.
 \] 
 On the other hand, there exists a factorisation of the lifing $f^{\sharp} = h_X \circ Tf$ in the category of Eilenberg-Moore algebras $\EM$:
 \[
 f^{\sharp} = (TY, \mu_Y) \overset{e_{\textnormal{im}(f^{\sharp})}}{\twoheadrightarrow} (\textnormal{im}(f^{\sharp}), h_{\textnormal{im}(f^{\sharp})}) \overset{m_{\textnormal{im}(f^{\sharp})}}{\hookrightarrow} (X,h_X).
 \]

 The next result shows that, up to isomorphism, the closure of the subobject $m_{\textnormal{im}(f)}$ with respect to the algebra $\mathbb{X}$ is given by the subobject $m_{\textnormal{im}(f^{\sharp})}$.
 
   \begin{restatable}{lemma}{closureofimage}
   \label{closureofimage}
 	$\overline{m_{\textnormal{\textnormal{im}}(f)}}^{\mathbb{X}} = m_{\textnormal{\textnormal{im}}(f^{\sharp})}$ in $\textnormal{Sub}(\mathbb{X})$. 	
   \end{restatable}

The following example shows that closing a minimal Moore automaton with additional algebraic structure can be realised by applying a monad of the type in \Cref{inducedmonad}.

\begin{example}[Closure of Minimal Moore Automata]
\label{closure_of_coalg}	
Let $F$ be the set endofunctor with $FX = B \times X^A$, for fixed sets $A$ and $B$. As $F$ preserves monomorphisms, the canonical epi-mono factorisation system of the category of sets lifts to the category $\textnormal{Coalg}(F)$, which consists of unpointed Moore automata with input $A$ and output $B$. 

For any language $L: A^* \rightarrow B$, there exists a size-minimal Moore automaton $\textnormal{M}_L$ that accepts $L$. It can be recovered as the epi-mono factorisation of the final $F$-coalgebra homomorphism $\textnormal{obs}: A^* \rightarrow  \Omega$, that is, $\textnormal{M}_L = m_{\textnormal{im}(\textnormal{obs})}$. In more detail, $\Omega$ is carried by $B^{A^*}$, \textnormal{obs} satisfies $\textnormal{obs}(w)(v) = L(wv)$, and $A^*$ is equipped with the $F$-coalgebra structure $\langle \varepsilon, \delta \rangle: A^* \rightarrow B \times (A^*)^A$ defined by $\varepsilon(w) = L(w)$ and $\delta(w)(a) = wa$ \cite{van2016master}. 

Any algebra structure $h: TB \rightarrow B$ over a set monad $T$ induces a canonical\footnote{\label{canonicaldistributivelawfootnote}Given an algebra $h: TB \rightarrow B$ for a set monad $T$, one can define a distributive law $\lambda$ between $T$ and $F$ with $FX = B \times X^A$ by $\lambda_X := (h \times \textnormal{st}) \circ \langle T\pi_1, T\pi_2 \rangle: TFX \rightarrow FTX$ \cite{jacobs2006bialgebraic}. (We write $\textnormal{st}$ for the usual strength function $\textnormal{st}: T(X^A) \rightarrow (TX)^A$ defined by $ \textnormal{st}(U)(a) = T(\textnormal{ev}_a)(U)$,
	where $\textnormal{ev}_{a}(f) = f(a)$.)} distributive law $\lambda$ between $T$ and $F$ with $FX = B \times X^A$.  It is well-known that $\lambda$-bialgebras are algebras over the monad $T_{\lambda}$ on $\textnormal{Coalg}(F)$ defined by $T_{\lambda}(X,k) = (TX, \lambda_X \circ Tk)$ and $T_{\lambda}f = Tf$ \cite{turi1997towards}. One such $T_{\lambda}$-algebra is the final $F$-coalgebra $\Omega$, when equipped with a canonical $T$-algebra structure induced by finality  \cite[Prop. 3]{jacobs2012trace}.

The functor $T_{\lambda}$ preserves epimorphisms in the category $\textnormal{Coalg}(F)$, if $T$ preserves epimorphisms in the category of sets. The latter is the case for every set functor. By \Cref{inducedmonad}, there thus exists a well-defined monad $\overline{(\cdot)}$ on $\textnormal{\textnormal{Sub}}(\Omega)$. 

By construction, the minimal Moore automaton $\textnormal{M}_L$ lives in $\textnormal{\textnormal{Sub}}(\Omega)$. Reviewing the constructions in \cite{zetzsche2021} shows that the minimal $\lambda$-bialgebra $\mathbb{M}_L$ for $L$ is given by the image of the lifting of $\textnormal{obs}$, that is, $\mathbb{M}_L = m_{\textnormal{im}(\textnormal{obs}^{\sharp})}$. From \Cref{closureofimage} it thus follows $\mathbb{M}_L = \overline{\textnormal{M}_L}$. In other words, the minimal $\lambda$-bialgebra for $L$ can be obtained from the minimal $F$-coalgebra for $L$ by closing the latter with respect to the $T_{\lambda}$-algebra structure of $\Omega$. 

For an example of the monad unit, observe how the minimal coalgebra in \Cref{m(l)} embeds into the minimal bialgebra in \Cref{overlineml}. 
\end{example}

The situation can be further generalised. We assume that i) $\mathscr{C}$ is a category with an $(\mathscr{E}, \mathscr{M})$-factorisation system; ii) $\lambda$ is a distributive law between a monad $T$ on $\mathscr{C}$ that preserves $\mathscr{E}$ and an endofunctor $F$ on $\mathscr{C}$ that preserves $\mathscr{M}$; iii) $(\Omega, h_{\Omega}, k_{\Omega})$ is a final $\lambda$-bialgebra.

\begin{restatable}{theorem}{coalgebratheorem}
\label{coalgebratheorem}
There exists a functor $\overline{(\cdot)}: \textnormal{Sub}(\Omega, k_{\Omega}) \rightarrow \textnormal{Sub}(\Omega, h_{\Omega}, k_{\Omega})$ that yields a monad on $\textnormal{Sub}(\Omega, k_{\Omega})$ and satisfies $\overline{m_{\textnormal{im}(\textnormal{obs}_{(X,k)})}} \cong m_{\textnormal{im}(\textnormal{obs}_{\textnormal{free}_T(X,k)})}$ in $\textnormal{Sub}(\Omega, h_{\Omega}, k_{\Omega})$, for any $F$-coalgebra $(X,k)$.
\end{restatable} 

To recover \Cref{closure_of_coalg} as a special case of \Cref{coalgebratheorem}, one instantiates the latter for the set endofunctor $F$ with $FX = B \times X^A$ and the canonical $F$-coalgebra with carrier $A^*$.

Finally, using analogous functors to the ones present in \Cref{closure_as_composition}, we observe that, as a consequence of \Cref{closureofimage}, the diagram in \Cref{closure_commuting} commutes.

\section{Step 2: Generators and Bases}

\label{generatorsandbases_sec}

One of the central notions of linear algebra is the \emph{basis}: a subset of a vector space is called basis, if every vector can be uniquely  written as a linear combination of basis elements. 

Part of the importance of bases stems from the convenient consequences that follow from their existence. For example, linear transformations between vector spaces admit matrix representations relative to pairs of bases \cite{lang2004algebra}, which can be used for efficient calculations. The idea of a basis however is not restricted to the theory of vector spaces: other algebraic theories have analogous notions of bases -- and generators, by waiving the uniqueness constraint --, for instance modules, semi-lattices, Boolean algebras, convex sets, and many more. In fact, the theory of bases for vector spaces is special only in the sense that every vector space admits a basis, which is not  the case for e.g. modules. 

In this section, we use the abstraction of generators and bases given in \Cref{generatordefinition} to lift results from one theory to the others. For example, one may wonder if there exists a matrix representation theory for convex sets that is analogous to the one of vector spaces.

\subsection{Categorification}
This section introduces a notion of morphism between algebras with a generator or a basis.

\begin{definition}%[$\textnormal{GAlg}(T)$]
\label{generatorcategory}
The category $\textnormal{GAlg}(T)$ of algebras with a generator over a monad $T$ is defined as: 
objects are pairs $(\mathbb{X}_{\alpha}, \alpha)$, where $\mathbb{X}_{\alpha} = (X_{\alpha}, h_{\alpha})$ is a $T$-algebra with generator $\alpha = (Y_{\alpha}, i_{\alpha}, d_{\alpha})$;
	 a morphism $(f, p): (\mathbb{X}_{\alpha}, \alpha) \rightarrow (\mathbb{X}_{\beta}, \beta)$ consists of a $T$-algebra homomorphism $f: \mathbb{X}_{\alpha} \rightarrow \mathbb{X}_{\beta}$ and a Kleisli-morphism $p: Y_{\alpha} \rightarrow TY_{\beta}$, such that the diagram below commutes:
	\begin{equation}
	\label{galgmorph}
	\begin{tikzcd}[row sep = 0.75em]
		X_{\alpha}  \arrow{r}{d_{\alpha}} \arrow{d}[left]{f} & TY_{\alpha} \arrow{d}{p^{\sharp}}   \arrow{r}{i_{\alpha}^{\sharp}}  & X_{\alpha} \arrow{d}{f} \\
		 X_{\beta}  \arrow{r}{d_{\beta}} & TY_{\beta} \arrow{r}{i_{\beta}^{\sharp}} &  X_{\beta} 
	\end{tikzcd}.
	\end{equation}
	Given $(f, p): (\mathbb{X}_{\alpha}, \alpha) \rightarrow (\mathbb{X}_{\beta}, \beta)$ and $(g, q): (\mathbb{X}_{\beta}, \beta) \rightarrow (\mathbb{X}_{\gamma}, \gamma)$, their composition is defined componentwise as $(g, q) \circ (f, p) := (g \circ f, q \cdot p)$, where $q \cdot p := \mu_{Y_{\gamma}} \circ Tq \circ p$ denotes the usual Kleisli-composition.
\end{definition}

The category $\textnormal{BAlg}(T)$ of algebras with a basis is defined as full subcategory of $\textnormal{GAlg}(T)$.

Let $F:  \EM \rightarrow \textnormal{GAlg}(T)$ be the functor with $F(\mathbb{X})  := (\mathbb{X}, (X, \textnormal{id}_X, \eta_{X}))$ and $F(f: \mathbb{X} \rightarrow \mathbb{Y}) := (f, \eta_Y \circ f)$, and $U: \textnormal{GAlg}(T) \rightarrow \EM$ the forgetful functor defined as the projection on the first component. Then $F$ and $U$ are in an adjoint relation:
\begin{restatable}{lemma}{galgfreeforgetful}
	\label{galgfreeforgetful}
$F \dashv U \,\colon\, \textnormal{GAlg}(T) \leftrightarrows \EM$.
\end{restatable}

\subsection{Products}

\label{productofbasessec}

In this section we show that, under certain assumptions, the monoidal product of a category naturally extends to a monoidal product of algebras with bases within that category. As a natural example we obtain the tensor-product of vector spaces with fixed bases.

We assume basic familiarity with monoidal categories. A monoidal monad $T$ on a monoidal category $(\mathscr{C}, \otimes, I)$ is a monad which is equipped with natural transformations $T_{X,Y}: TX \otimes TY \rightarrow T(X \otimes Y)$ and $T_0: I \rightarrow TI$, satisfying certain coherence conditions (see e.g. \cite{seal2013tensors}). One can show that, given such additional data, the monoidal structure of $\mathscr{C}$ induces a monoidal category $(\EM, \boxtimes, (TI, \mu_I))$, if two appropriately defined\footnote{
	(A1) For any two algebras $\mathbb{X}_{\alpha} = (X_{\alpha}, h_{\alpha})$ and $\mathbb{X}_{\beta} = (X_{\beta}, h_{\beta})$ the coequaliser $q_{\mathbb{X}_{\alpha}, \mathbb{X}_{\beta}}$ of the algebra homomorphisms 
$T(h_{\alpha} \otimes h_{\beta})$ and $\mu_{X_{\alpha} \otimes X_{\beta}} \circ T(T_{X_{\alpha}, X_{\beta}})$ of type $(T(TX_{\alpha} \otimes TX_{\beta}), \mu_{TX_{\alpha} \otimes TX_{\beta}}) \rightarrow (T(X_{\alpha} \otimes X_{\beta}), \mu_{X_{\alpha} \otimes X_{\beta}})$ exists (we denote its codomain by $\mathbb{X}_{\alpha} \boxtimes \mathbb{X}_{\beta} := (X_{\alpha} \boxtimes X_{\beta}, h_{\alpha \boxtimes \beta})$).
(A2) Left and right-tensoring with the induced functor $\boxtimes$ preserves reflexive coequalisers.
}
assumptions (A1) and (A2) are satisfied
\cite[Cor. 2.5.6]{seal2013tensors}.
The two monoidal products $\otimes$ and $\boxtimes$ are related via the natural embedding $q_{\mathbb{X}_{\alpha}, \mathbb{X}_{\beta}} \circ \eta_{X_{\alpha} \otimes X_{\beta}}$, in the following referred to as $\iota_{\mathbb{X}_{\alpha}, \mathbb{X}_{\beta}}$. One can prove that the product $TY_{\alpha} \boxtimes TY_{\beta}$ is given by $T(Y_{\alpha} \otimes Y_{\beta})$ and the coequaliser $q_{TY_{\alpha}, TY_{\beta}}$ by $\mu_{Y_{\alpha} \otimes Y_{\beta}} \circ T(T_{Y_{\alpha}, Y_{\beta}})$, where we abbreviate the free algebra $(TY, \mu_Y)$ as $TY$ \cite{seal2013tensors}. 

With the previous remarks in mind, we are able to claim the following.
%From the universality of a coequaliser can can deduce the existence of a (unique) algebra homomorphism $T^{\boxtimes}_{Y_{\alpha}, Y_{\beta}}: (TY_{\alpha}, \mu_{Y_{\alpha}}) \boxtimes (TY_{\beta}, \mu_{Y_{\beta}}) \rightarrow (T(Y_{\alpha} \otimes Y_{\beta}), \mu_{Y_{\alpha} \otimes Y_{\beta}})$.

\begin{restatable}{lemma}{productofbases}
\label{productofbases}
	Let $T$ be a monoidal monad on $(\mathscr{C}, \otimes, I)$ satisfying (A1) and (A2). Let $\alpha = (Y_{\alpha}, i_{\alpha}, d_{\alpha})$ and $\beta = (Y_{\beta}, i_{\beta}, d_{\beta})$ be generators (bases) for $T$-algebras $\mathbb{X}_{\alpha}$ and $\mathbb{X}_{\beta}$. Then $\alpha \boxtimes \beta = (Y_{\alpha} \otimes Y_{\beta}, \iota_{\mathbb{X}_{\alpha}, \mathbb{X}_{\beta}} \circ (i_{\alpha} \otimes i_{\beta}), (d_{\alpha} \boxtimes d_{\beta}))$ is a generator (basis) for the $T$-algebra $\mathbb{X}_{\alpha} \boxtimes \mathbb{X}_{\beta}$.		
\end{restatable}

\begin{restatable}{corollary}{monoidalproduct}
\label{monoidalproduct}
Let $T$ be a monoidal monad on $(\mathscr{C}, \otimes, I)$ such that (A1) and (A2) are satisfied. The definitions $(\mathbb{X}_{\alpha}, \alpha) \boxtimes (\mathbb{X}_{\beta}, \beta) := (\mathbb{X}_{\alpha} \boxtimes \mathbb{X}_{\beta}, \alpha \boxtimes \beta)$ and $(f, p) \boxtimes (g,q) := (f \boxtimes g, T_{Y_{\alpha'}, Y_{\beta'}} \circ (p \otimes q))$ yield monoidal structures with unit $((TI, \mu_I), (I, \eta_I, \textnormal{id}_{TI}))$ on $\textnormal{GAlg}(T)$ and $\textnormal{BAlg}(T)$. 	
\end{restatable}

We conclude by instantiating above construction to the setting of vector spaces.

\begin{example}[Tensor Product of Vector Spaces]
	Recall the free $\mathbb{K}$-vector space monad $\mathcal{V}_{\mathbb{K}}$ defined by $\mathcal{V}_{\mathbb{K}}(X) = X \rightarrow \mathbb{K}$ and  $\mathcal{V}_{\mathbb{K}}(\varphi)(y) = \sum_{x \in f^{-1}(y)} \varphi(x)$.  Its unit is given by $\eta_X(x)(y) = \lbrack x = y \rbrack$ and its multiplication by $\mu_X(\Phi)(x) = \sum_{\varphi \in {\mathbb{K}}^X} \Phi(\varphi) \cdot \varphi(x)$. 
	
	The category of sets is monoidal (in fact, cartesian) with respect to the cartesian product $\times$ and the singleton set $\lbrace * \rbrace$. The monad $\mathcal{V}_{\mathbb{K}}$ is monoidal when equipped with $(\mathcal{V}_{\mathbb{K}})_{X,Y}(\varphi, \psi)(x,y) := \varphi(x) \cdot \psi(y)$ and $(\mathcal{V}_{\mathbb{K}})_0(*)(*) := 1_{\mathbb{K}}$ \cite{parlant_et_al:LIPIcs:2020:12746}. The category of $\mathcal{V}_{\mathbb{K}}$-algebras is isomorphic to the category of $\mathbb{K}$-vector spaces, and satisfies (A1) and (A2). The monoidal structure induced by $\mathcal{V}_{\mathbb{K}}$ is the usual tensor product $\otimes$ with the unit field $\mathcal{V}_{\mathbb{K}}(\lbrace * \rbrace) \simeq \mathbb{K}$. 
	
	\Cref{productofbases} captures the well-known fact that the dimension of the tensor product of two vector spaces is the product of the respective dimensions. The structure maps of the product generator map $(y_{\alpha}, y_{\beta})$ to the vector $i(y_{\alpha}) \otimes i(y_{\beta})$, and $x$ to $(d_{\alpha} \otimes d_{\beta})(x)$, where 
	\begin{align*}
		d_{\alpha} \otimes d_{\beta} = \overline{d_{\alpha} \times d_{\beta}}: \mathbb{X}_{\alpha} \otimes \mathbb{X}_{\beta} \rightarrow (\mathcal{V}_{\mathbb{K}}(Y_{\alpha}), \mu_{Y_{\alpha}}) \otimes (\mathcal{V}_{\mathbb{K}}(Y_{\beta}), \mu_{Y_{\beta}}) \simeq (\mathcal{V}_{\mathbb{K}}(Y_{\alpha} \times Y_{\beta}), \mu_{Y_{\alpha} \otimes Y_{\beta}})
	\end{align*}  is the unique linear extension of the bilinear map defined by \[(d_{\alpha} \times d_{\beta})(x_{\alpha}, x_{\beta})(y_{\alpha}, y_{\beta}) := d_{\alpha}(x_{\alpha})(y_{\alpha}) \cdot d_{\beta}(x_{\beta})(y_{\beta}). \] 
\end{example}

\subsection{Kleisli Representation Theory}

\label{representationtheorysec}

In this section we use our category-theoretical definition of a basis to derive a representation theory for homomorphisms between algebras over monads that is analogous to the matrix representation theory for linear transformations between vector spaces.

In more detail, recall that a linear transformation $L: V \rightarrow W$ between $k$-vector spaces with finite bases $\alpha = \lbrace v_1, ... , v_n \rbrace$ and $\beta = \lbrace w_1,  ..., w_m \rbrace$, respectively, admits a matrix representation $L_{\alpha \beta} \in \textnormal{Mat}_{k}(m, n)$ with  $ L(v_j) = \sum_i (L_{\alpha \beta})_{i,j} w_i$, such that for any vector $v$ in $V$ the coordinate vectors $L(v)_{\beta} \in k^m$ and $v_{\alpha} \in k^n$   
satisfy the equality $L(v)_{\beta} = L_{\alpha \beta} v_{\alpha}$. A great amount of linear algebra is concerned with finding bases such that the corresponding matrix representation is in an efficient shape, for instance diagonalised. The following definitions generalise the situation by substituting Kleisli morphisms for matrices. 
  
\begin{definition}
\label{representationdef}
	Let $\alpha = (Y_{\alpha}, i_{\alpha}, d_{\alpha})$ and $\beta = (Y_{\beta}, i_{\beta}, d_{\beta})$ be bases for $T$-algebras $\mathbb{X}_{\alpha} = (X_{\alpha},h_{\alpha})$ and $\mathbb{X}_{\beta} = (X_{\beta},h_{\beta})$, respectively. The basis representation $f_{\alpha \beta}$ of a $T$-algebra homomorphism $f: \mathbb{X}_{\alpha} \rightarrow \mathbb{X}_{\beta}$ with respect to $\alpha$ and $\beta$ is defined by 
	\begin{equation}
	\label{basisrepresentation}
		f_{\alpha \beta} := Y_{\alpha} \overset{i_{\alpha}}{\longrightarrow} X_{\alpha} \overset{f}{\longrightarrow} X_{\beta} \overset{d_{\beta}}{\longrightarrow} TY_{\beta}.
	\end{equation}
	 Conversely, the morphism $p^{\alpha \beta}$ associated with a Kleisli morphism $p: Y_{\alpha} \rightarrow TY_{\beta}$ with respect to $\alpha$ and $\beta$ is defined by \begin{equation}
		\label{associatedmorph}
		p^{\alpha \beta} := X_{\alpha} \overset{d_{\alpha}}{\longrightarrow} TY_{\alpha} \overset{Tp}{\longrightarrow} T^2Y_{\beta} \overset{\mu_{Y_{\beta}}}{\longrightarrow} TY_{\beta} \overset{Ti_{\beta}}{\longrightarrow} TX_{\beta} \overset{h_{\beta}}{\longrightarrow} X_{\beta}.
			\end{equation}
	\end{definition}	
	
	\begin{figure*}[t]
\[ A = L_{\alpha' \alpha'} = 
\begin{pmatrix}
	  0 & -1 \\ 1 & 0
\end{pmatrix}, 
\quad
L_{\alpha \alpha} = \begin{pmatrix}
	3 & 2 \\
	-5 & -3
\end{pmatrix},
\quad
P = \begin{pmatrix}
	-1 & 1 \\
	2 & -1
\end{pmatrix},
\quad
P^{-1} = \begin{pmatrix}
	1 & 1 \\ 2 & 1
\end{pmatrix}
\]
	\caption{The basis representation of the counter-clockwise rotation by 90 degree $L: \mathbb{R}^2 \rightarrow \mathbb{R}^2$, $L(v) = Av$ with respect to $\alpha = \lbrace (1,2), (1,1) \rbrace$ and $\alpha' = \lbrace (1,0), (0,1) \rbrace$ satisfies $L_{\alpha' \alpha'} = P^{-1} L_{\alpha \alpha} P$.}
		\label{basisrepex1}
\end{figure*}

The associated morphism is the linear transformation between vector spaces induced by some matrix of the right type. The following result confirms this intuition.

\begin{restatable}{lemma}{associsalgebrahom}
\label{associsalgebrahom}
	The function \eqref{associatedmorph} is a $T$-algebra homomorphism $p^{\alpha \beta}: \mathbb{X}_{\alpha} \rightarrow \mathbb{X}_{\beta}$.	
\end{restatable}

The next result establishes a generalisation of the observation that for fixed bases, constructing a matrix representation of a linear transformation and associating a linear transformation to a matrix of the right type are mutually inverse operations.

\begin{restatable}{lemma}{inverseoperations}
\label{inverseoperations}
	The operations $\eqref{basisrepresentation}$ and $\eqref{associatedmorph}$ are mutually inverse.		
\end{restatable}

At the beginning of this section we recalled the soundness identity $L(v)_{\beta} = L_{\alpha \beta} v_{\alpha}$ for the matrix representation $L_{\alpha \beta}$ of a linear transformation $L$. The next result is a natural generalisation of this statement.

\begin{restatable}{lemma}{uniquehom}
	\label{uniquehom}
	$f_{\alpha \beta}$ is the unique Kleisli-morphism such that $f_{\alpha \beta} \cdot d_{\alpha} = d_{\beta} \circ f $. Conversely, $p^{\alpha \beta}$ is the unique $T$-algebra homomorphism such that $p \cdot d_{\alpha} = d_{\beta} \circ p^{\alpha \beta}$.  	
\end{restatable}

The next result establishes the compositionality of the operations $\eqref{basisrepresentation}$ and $\eqref{associatedmorph}$. For example, the matrix representation of the composition of two linear transformations is given by the multiplication of the matrix representations of the individual linear transformations.

\begin{restatable}{lemma}{compboth}
\label{compboth}
$g_{\beta \gamma} \cdot f_{\alpha \beta} = (g \circ f)_{\alpha \gamma}$ and $q^{\beta \gamma} \circ p^{\alpha \beta} = (q \cdot p)^{\alpha \gamma}$.
\end{restatable}

The previous statements may be summarised as functors between appropriately defined\footnote{Let $\textnormal{Alg}_{\textnormal{B}}(T)$ be the category in which objects are given by pairs $(\mathbb{X}_{\alpha}, \alpha)$, where $\mathbb{X}_{\alpha}$ is a $T$-algebra with basis $\alpha = (Y_{\alpha}, i_{\alpha}, d_{\alpha})$, and a morphism $f: (\mathbb{X}_{\alpha}, \alpha) \rightarrow (\mathbb{X}_{\beta}, \beta)$ consists of a $T$-algebra homomorphism $f: \mathbb{X}_{\alpha} \rightarrow \mathbb{X}_{\beta}$.
Let $\textnormal{Kl}_{\textnormal{B}}(T)$ be the category in which objects are the same ones as for $\textnormal{Alg}_{\textnormal{B}}(T)$, and a morphism $p: (\mathbb{X}_{\alpha}, \alpha) \rightarrow (\mathbb{X}_{\beta}, \beta)$ consists of a Kleisli-morphism $p: Y_{\alpha} \rightarrow TY_{\beta}$.} categories $\textnormal{Alg}_{\textnormal{B}}(T)$ and $\textnormal{Kl}_{\textnormal{B}}(T)$.

\begin{restatable}{corollary}{isomorphismrep}
\label{isomorphismrep}
	There exist isomorphisms of categories
	$\textnormal{BAlg}(T) \simeq \textnormal{Alg}_{\textnormal{B}}(T) \simeq \textnormal{Kl}_{\textnormal{B}}(T)$.	
\end{restatable}

Assume we are given bases $\alpha, \alpha'$ and $\beta, \beta'$ for $T$-algebras $(X_{\alpha}, h_{\alpha})$ and $(X_{\beta}, h_{\beta})$, respectively. 
The following result clarifies how the representations $f_{\alpha \beta}$ and $f_{\alpha' \beta'}$ are related.

\begin{restatable}{proposition}{similaritygeneral}
\label{similaritygeneral}
	There exist Kleisli isomorphisms $p$ and $q$ such that $f_{\alpha' \beta'} = q \cdot f_{\alpha \beta} \cdot p$.	
\end{restatable}

Above result simplifies if one restricts to an endomorphism: the basis representations are \emph{similar}. This generalises the situation for vector spaces, cf. \Cref{basisrepex1}.

\begin{restatable}{proposition}{similarity}
		\label{similarity}
There exists a Kleisli isomorphism $p$ with Kleisli inverse $p^{-1}$ such that $f_{\alpha' \alpha'} = p^{-1} \cdot f_{\alpha \alpha} \cdot p$.
\end{restatable}

\subsection{Bases for Bialgebras}

\label{basesforbialgebrasec}

This section is concerned with generators and bases for \emph{bialgebras}. It is well-known \cite{turi1997towards} that an Eilenberg-Moore law $\lambda$ between a monad $T$ and an endofunctor $F$ induces simultaneously i) a monad $T_{\lambda} = (T_{\lambda}, \mu, \eta)$ on $\textnormal{Coalg}(F)$ by $T_{\lambda}(X,k) = (TX, \lambda_X \circ Tk)$ and $T_{\lambda}f = Tf$; and ii) an endofunctor $F_{\lambda}$ on $\EM$ by $F_{\lambda}(X,h) = (FX, Fh \circ \lambda_X)$ and $F_{\lambda}f = Ff$,
such that the algebras over $T_{\lambda}$, the coalgebras of $F_{\lambda}$, and $\lambda$-bialgebras coincide.
 We will consider generators and bases for $T_{\lambda}$-algebras, or equivalently, $\lambda$-bialgebras.

By \Cref{generatordefinition}, a generator for a $\lambda$-bialgebra $(X,h,k)$ consists of a $F$-coalgebra $(Y,k_Y)$ and morphisms $i: Y \rightarrow X$, $d: X \rightarrow TY$, such that the three squares on the left of \eqref{bialgebrageneratorequations} commute:
\begin{equation}
\label{bialgebrageneratorequations}	
	\begin{tikzcd}[row sep = 0.75em, column sep = 1.5em]
			Y \arrow{r}{i} \arrow{d}[left]{k_Y} & X \arrow{d}{k} \\
		FY \arrow{r}{Fi} &FX
	\end{tikzcd}
	\quad
		\begin{tikzcd}[row sep = 0.75em, column sep = 1.5em]
		X \arrow{r}{d} \arrow{d}[left]{k} & TY \arrow{d}[right ]{\lambda_Y \circ Tk_Y} \\
		FX \arrow{r}{Fd} &FTY
	\end{tikzcd}
	\quad
	\begin{tikzcd}[row sep = 0.75em, column sep = 1.5em]			TY \arrow{r}{Ti} & TX \arrow{d}{h}\\
			X \arrow{u}{d} \arrow{r}{\textnormal{id}_X}  & X 
		\end{tikzcd}
		\quad
		\begin{tikzcd}[row sep = 0.75em,column sep = 1.5em]
			TX \arrow{r}{h} & X \arrow{d}{d}\\
			TY \arrow{u}{Ti} \arrow{r}{\textnormal{id}_{TY}}  & TY 
		\end{tikzcd}.
\end{equation}

A basis for a bialgebra is a generator such that the diagram on the right of \eqref{bialgebrageneratorequations} commutes.

By forgetting the $F$-coalgebra structure, every generator for a bialgebra is in particular a generator for its underlying $T$-algebra. By \Cref{forgenerator-isharp-is-bialgebra-hom} there exists a $\lambda$-bialgebra homomorphism 
$ i^{\sharp} := h \circ Ti: \textnormal{exp}_T(Y, Fd \circ k \circ i) \rightarrow (X,h,k)$. The next result establishes that there exists a second equivalent free bialgebra with a different coalgebra structure.

\begin{restatable}{lemma}{generatorbialgebraisharp}
		Let $(Y,k_Y, i,d)$ be a generator for $(X,h,k)$. Then $i^{\sharp}: TY \rightarrow X$ is a $\lambda$-bialgebra homomorphism $i^{\sharp} : \textnormal{free}_T(Y, k_Y) \rightarrow (X,h,k)$.
\end{restatable}

If one moves from generators to bases for bialgebras, both structures coincide.

\begin{restatable}{lemma}{samecoalgebrastructures}
	\label{samecoalgebrastructures}
	Let $(Y,k_Y, i, d)$ be a basis for $(X,h,k)$, then $\textnormal{free}_T(Y, k_Y) = \textnormal{exp}_T(Y, Fd \circ k \circ i)$.
\end{restatable}

\begin{example}[Canonical RFSA]
Recall the minimal pointed bialgebra $(X, h, k)$ for the language $L= (a+b)^*a$ depicted in \Cref{overlineml}. Let $(J(\mathbb{X}), i, d)$ be the generator for $\mathbb{X} = (X,h)$ defined as follows: the carrier $J(\mathbb{X})$ consists of the join-irreducibles for $\mathbb{X}$, the embedding satisfies $i(y) = y$, and the decomposition is given by $d(x) = \lbrace y \in J(\mathbb{X}) \mid y \leq x \rbrace$). We used $(J(\mathbb{X}), i, d)$ to recover the canonical RFSA for $L$ depicted in \Cref{jiromaton} as the coalgebra $(J(\mathbb{X}), Fd \circ k \circ i)$. Examining the graphs shows that $k$ restricts to the join-irreducibles $J(\mathbb{X})$, suggesting $\alpha = (J(\mathbb{X}), k, i, d)$ as a possible generator for the full bialgebra. However, the $a$-action on $\lbrack \lbrace y \rbrace \rbrack$ implies the non-commutativity of the second diagram on the left of \eqref{bialgebrageneratorequations}. The issue can be fixed by modifying $d$ via $d(\lbrack \lbrace y \rbrace \rbrack) := \lbrace \lbrack \lbrace y \rbrace \rbrack \rbrace$. In consequence $\textnormal{free}(J(\mathbb{X})), k)$ and $\textnormal{exp}(J(\mathbb{X}), Fd \circ k \circ i)$ coincide (even though the assumptions of \Cref{samecoalgebrastructures} are not satisfied).	\end{example}

We close this section by observing that a basis for the underlying algebra of a bialgebra is sufficient for constructing a generator for the full bialgebra.

\begin{restatable}{lemma}{generatorfullbialgebra}
		\label{generatorfullbialgebra}
	Let $(X,h,k)$ be a $\lambda$-bialgebra and $(Y,i,d)$ a basis for the $T$-algebra $(X,h)$. Then $(TY,F\mu_Y \circ \lambda_{TY} \circ T(Fd \circ k \circ i)
,h \circ Ti,\eta_{TY} \circ d)$ is a generator for $(X,h,k)$.
\end{restatable}

\subsection{Bases as Coalgebras}

\label{basesascoaglebrassec}

In this section, we compare our approach to an alternative perspective on the generalisation of bases. 
More specifically, we are interested in the work of Jacobs \cite{jacobs2011bases}, where a basis is defined as a coalgebra for the comonad on the category of Eilenberg-Moore algebras induced by the free algebra adjunction. Explicitly, a basis for a $T$-algebra $(X,h)$, in the sense of \cite{jacobs2011bases}, consists of a $T$-coalgebra $(X,k)$ such that the following three diagrams commute:
\begin{equation}
\label{jacobsbasis}
\begin{tikzcd}[row sep = 0.75em,]
			TX \arrow{d}[left]{h} \arrow{r}{Tk} & T^2X \arrow{d}{\mu_X} \\
			X \arrow{r}{k} & TX
		\end{tikzcd}
		\qquad
	\begin{tikzcd}[row sep = 0.75em,]			X \arrow{dr}[left]{\textnormal{id}_X} \arrow{r}{k} & TX \arrow{d}{h} \\
			& X
		\end{tikzcd}	
		\qquad
		\begin{tikzcd}[row sep = 0.75em,]
			X \arrow{d}[left]{k} \arrow{r}{k} & TX \arrow{d}{T\eta_X} \\
			TX \arrow{r}{Tk} & T^2X
		\end{tikzcd}.
\end{equation}

The next result shows that a basis as in \Cref{generatordefinition} induces a basis in the sense of \cite{jacobs2011bases}.

\begin{restatable}{lemma}{impliesjacobsbasis}
\label{impliesjacobsbasis}
	Let $(Y,i,d)$ be a basis for a $T$-algebra $(X,h)$, then \eqref{jacobsbasis} commutes for $k := Ti \circ d$.	
\end{restatable}

Conversely, assume $(X,k)$ is a $T$-coalgebra structure satisfying \eqref{jacobsbasis} and $i_k: Y_k \rightarrow X$ an equaliser  of $k$ and $\eta_X$. If the underlying category is the usual category of sets, the equaliser of any two functions exists. If $Y_k$ non-empty, one can show that the equaliser is preserved under $T$, that is, $Ti_k$ is an equaliser of $Tk$ and $T\eta_X$ \cite{jacobs2011bases}. By \eqref{jacobsbasis} we have $Tk \circ k = T\eta_X \circ k$. Thus there exists a unique morphism $d_k: X \rightarrow TY_k$ such that $Ti_k \circ d_k = k$, which can be shown to be the inverse of $h \circ Ti_k$ \cite{jacobs2011bases}. In other words, $G(X,k) := (Y_k, i_k, d_k)$ is a basis for $(X,h)$ in the sense of \Cref{generatordefinition}. In the following let $F(Y,i,d) := (X,Ti \circ d)$ for an arbitrary basis of $(X,h)$.

\begin{restatable}{lemma}{equaliserlemma}
	\label{equaliserlemma}	
	Let $(Y,i,d)$ be a basis for a $T$-algebra $(X,h)$ and $k := Ti \circ d$. Then $\eta_X \circ i = k \circ i$ and $Tk \circ (\eta_X \circ i) = T\eta_{X} \circ (\eta_X \circ i)$.  
\end{restatable}

\begin{restatable}{corollary}{uniquemorphism}
\label{uniquemorphism}
	Let $\alpha := (Y,i,d)$  be a basis for a set-based $T$-algebra $(X,h)$ and $k := Ti \circ d$. Let $i_k: Y_k \rightarrow X$ be an equaliser of $k$ and $\eta_X$, and $Y_k$ non-empty, then $(\textnormal{id}_{(X,h)})_{\alpha, GF\alpha}: Y \rightarrow TY_{k}$ is the unique morphism $\psi$ making the diagram below commute:
\[
\begin{tikzcd}[ampersand replacement=\&, row sep = 0.75em]
Y \arrow[dashed]{r}[below]{\psi} \arrow[bend left=20]{rr}{\eta_X \circ i} \& TY_k \arrow{r}[below]{Ti_k}	 \& TX \arrow[shift left=2]{r}[above]{Tk} \arrow[shift right=1.5]{r}[below]{T\eta_X} \& T^2X
\end{tikzcd}.
\]	
\end{restatable}

\subsection{Signatures, Equations, and Finitary Monads}

\label{varietiessec}

Most of the algebras over set monads one usually considers generators for constitute finitary varieties in the sense of universal algebra. In this section, we will briefly explore the consequences for generators that arise from this observation. The constructions are well-known; we include them for completeness.

 Let $\Sigma$ be a set, whose elements we think of as \emph{operations}, and $\textnormal{ar}: \Sigma \rightarrow \mathbb{N}$ a function that assigns to an operation its \emph{arity}. Any such \emph{signature} induces a set endofunctor $H_{\Sigma}$ defined on a set as the coproduct $H_{\Sigma}X = \coprod_{\sigma \in \Sigma} X^{\textnormal{ar}(\sigma)}$, and consequently, a set monad $\mathbb{S}_{\Sigma}$ that assigns to a set $V$ of variables the initial algebra $S_{\Sigma} V = \mu X.(V + H_{\Sigma} X)$, i.e. the set of $\Sigma$-terms generated by $V$ (see e.g. \cite{turi1996functorial}). One can show that the categories of $H_{\Sigma}$-algebras and $\mathbb{S}_{\Sigma}$-algebras are isomorphic. A $\mathbb{S}_{\Sigma}$-algebra $\mathbb{X}$ \emph{satisfies a set of equations} $E \subseteq S_{\Sigma} V \times S_{\Sigma} V$, if for all $(s,t) \in E$ and valuations $v: V \rightarrow X$ it holds $v^{\sharp}(s) = v^{\sharp}(t)$, where $v^{\sharp}: (S_{\Sigma}V, \mu_V) \rightarrow \mathbb{X}$ is the unique extension of $v$ to a $\mathbb{S}_{\Sigma}$-algebra homomorphism \cite{adamek1994locally}. The set of $\mathbb{S}_{\Sigma}$-algebras that satisfy $E$ is denoted by $\textnormal{Alg}(\Sigma, E)$. As one verifies, the forgetful functor $U: \textnormal{Alg}(\Sigma, E) \rightarrow \Set$ admits a left-adjoint $F: \Set \rightarrow \textnormal{Alg}(\Sigma, E)$, thus resulting in a set monad $T_{\Sigma, E}$ with underlying endofunctor $U \circ F$ that preserves directed colimits. The functor $U$ can be shown to be monadic, that is, the comparison functor $K: \textnormal{Alg}(\Sigma, E) \rightarrow \textnormal{Set}^{T_{\Sigma, E}}$ is an isomorphism \cite{mac2013categories}. In other words, the category of Eilenberg-Moore algebras over $T_{\Sigma, E}$ and the finitary variety of algebras over $\Sigma$ and $E$ coincide. In fact, set monads preserving directed colimits (sometimes called \emph{finitary} monads \cite{adamek1994locally}) and finitary varieties are in \emph{bijection}.
 
The following result characterises generators for algebras over $T_{\Sigma, E}$. It can be seen as a unifying proof for observations analogous to the one in \Cref{setbasedexamples}. 

\begin{restatable}{lemma}{sigmatermlemma}
\label{sigmatermlemma}
A morphism $i: Y \rightarrow X$ is part of a generator for a $T_{\Sigma, E}$-algebra $\mathbb{X}$ iff every element $x \in X$ can be expressed as a $\Sigma$-term in $i\lbrack Y \rbrack$ modulo $E$, that is, there is a term $d(x) \in S_{\Sigma}Y$ such that $i^{\sharp}(\llbracket d(x) \rrbracket_E) = x$.\end{restatable}

\subsection{Finitely Generated Objects}

\label{finitelygeneratedsec}

In this section, we relate our abstract definition of a generator to the theory of \emph{locally finitely presentable} categories, in particular, to the notions of \emph{finitely generated} and \emph{finitely presentable} objects, which are categorical abstractions of finitely generated algebraic structures.

For intuition, recall that an element $x \in X$ of a partially ordered set is \emph{compact}, if for each directed set $D \subseteq X$ with $x \leq \bigvee D$, there exists some $d \in D$ satisfying $x \leq d$. An \emph{algebraic lattice} is a partially ordered set that has all joins, and every element is a join of compact elements. The naive categorification of compact elements is equivalent to the following definition: a object $Y$ in $\mathscr{C}$ is \emph{finitely presentable (generated)}, if $\textnormal{Hom}_{\mathscr{C}}(Y, -): \mathscr{C} \rightarrow \Set$ preserves filtered colimits (of monomorphisms). Consequently, one can categorify algebraic lattices as  \emph{locally finitely presentable} (lfp) categories, which are cocomplete and admit a set of finitely presentable objects, such that every object is a filtered colimit of that set \cite{adamek1994locally}.

 In \cite[Theor. 3.5]{adamek2019finitely} it is shown that an algebra $\mathbb{X}$ over a finitary monad $T$ on an lfp category $\mathscr{C}$ is a finitely generated object of $\EM$ iff there exists a finitely presentable object $Y$ of $\mathscr{C}$ and a morphism $i: Y \rightarrow X$, such that $i^{\sharp}: (TY, \mu_Y) \rightarrow \mathbb{X}$ is a strong\footnote{An epimorphism $e: A \rightarrow B$ is said to be \emph{strong}, if for any monomorphism $m: C \rightarrow D$ and any morphisms $f: A \rightarrow C$ and $g: B \rightarrow D$ such that $g \circ e = m \circ f$, there exists a diagonal monomorphism $d: B \rightarrow C$ such that $f = d \circ e$ and $g = m \circ d$.} epimorphism in $\EM$. Below, we give a variant of this statement where instead the carrier of $i^{\sharp}$ is a split\footnote{A morphism $e: A \rightarrow B$ is called \emph{split}, if there exists a morphism $s: B \rightarrow A$ such that $e \circ s = \textnormal{id}_B$. Any morphism that is split is necessarily a strong epimorphism.} epimorphism in $\mathscr{C}$, which is the case iff $\mathbb{X}$ admits a generator in the sense of \Cref{generatordefinition}.
  
 \begin{restatable}{proposition}{finitarymonadlemma}
 Let $\mathscr{C}$ be a lfp category in which strong epimorphisms split and $T$ a finitary monad on $\mathscr{C}$ preserving epimorphisms.
 Then an algebra $\mathbb{X}$ over $T$ is a finitely generated object of $\EM$ iff it is generated by a finitely presentable object $Y$ in $\mathscr{C}$ in the sense of \Cref{generatordefinition}.
 \end{restatable}

\section{Related Work}

\label{relatedworksec_closure}

A central motivation for this paper has been our broad interest in active learning algorithms for state-based models \cite{angluin1987learning}.
One of the challenges in learning non-deterministic models is the common lack of a unique minimal acceptor for a given language \cite{denis2001residual}. The problem has been independently approached for different variants of non-determinism, often with the common idea of finding a subclass admitting a unique representative \cite{esposito2002learning, berndt2017learning}. Unifying perspectives were given by van Heerdt  \cite{van2020learning, van2016master, van2020phd} and Myers et al. \cite{MyersAMU15}. One of the central notions in the work of van Heerdt is the concept of a scoop, originally introduced by Arbib and Manes \cite{arbib1975fuzzy}.

 In \cite{zetzsche2021} we have presented a categorical framework that recovers minimal non-deterministic representatives in two steps. The framework is based on ideas closely related to the ones in \cite{MyersAMU15}, adopts scoops under the name generators, and strengthens the former to the notion of a basis. In a first step, it constructs a minimal bialgebra by closing a minimal coalgebra with additional algebraic structure over a monad. In a second step, it identifies generators for the algebraic part of the bialgebra, to derive an equivalent coalgebra with side effects in a monad. 
  In this paper, we generalise the first step as application of a monad on an appropriate category of subobjects with respect to a $(\mathscr{E}, \mathscr{M})$-factorisation system, and explore the second step by further developing the abstract theory of generators and bases.
  
  	Categorical factorisation systems are well-established \cite{bousfield1977constructions,riehl2008factorization,maclane1950duality}. Among others, they have been used for a general view on the minimisation and determinisation of state-based systems \cite{adamek2009abstract, adamek2012coalgebraic, wissmann2022minimality}. In \Cref{closure_sec} we use the formalism of \cite{adamek2009abstract}. In \Cref{factorisationsystemalgebra} we have shown that under certain assumptions factorisation systems can be lifted to the categories of algebras and coalgebras. We later realised that the constructions had recently been published in \cite{wissmann2022minimality}.
  
The notion of a basis for an algebra over an arbitrary monad has been subject of previous interest. Jacobs, for instance, defines a basis as a coalgebra for the comonad on the category of algebras induced by the free algebra adjunction \cite{jacobs2011bases}. In \Cref{basesascoaglebrassec} we show that a basis in our sense always induces a basis in their sense, and, conversely, it is possible to recover a basis in our sense from a basis in their sense, if certain assumptions about the existence and preservation of equalisers are given. As equalisers do not necessarily exist and are not necessarily preserved, our approach carries additional data and thus can be seen as finer.

\section{Discussion and Future Work}

We have generalised the closure of a subset of an algebraic structure as a monad between categories of subobjects relative to a factorisation system. We have identified the closure of a minimal coalgebra as an instance of the closure of subobjects that arise by taking the image of a morphism. We have extended the notion of a generator to a category of algebras with generators, and explored its characteristics.
  We have generalised the matrix representation theory of vector spaces and discussed bases for bialgebras. We compared our ideas with a coalgebraic generalisation of bases, explored the case in which a monad is induced by a variety, and related our notion to finitely generated objects in finitely presentable categories. 
  
In \cite{zetzsche2021} we have shown that generators and bases in our sense are central ingredients in the definitions of minimal canonical acceptors. Many such acceptors admit double-reversal characterisations \cite{brzozowski1962canonical,BrzozowskiT14,MyersAMU15,VuilleminG210}. Duality based characterisations as the former have been shown to be closely related to minimisation procedures with respect to factorisation systems \cite{bonchi2012brzozowski,bonchi2014algebra,wissmann2022minimality}. In the future, it would be interesting to further explore the connection between the minimality of generators on the one side, and the minimality of an acceptor with respect to a factorisation system on the other side.

Another interesting question is whether the construction that underlies our definition of a monad in \Cref{inducedmonad} could be introduced at a more general level of an arbitrary adjunction between categories with suitable factorisation systems, such that the adjunction between the base category $\mathscr{C}$ and the category of Eilenberg-Moore algebras $\EM$ is a special case.

%%
%% Bibliography
%%

%% Please use bibtex, 

\bibliography{literature.bib}

\begin{thebibliography}{10}

\bibitem{adamek2012coalgebraic}
Jiri Adamek, Filippo Bonchi, Mathias H{\"u}lsbusch, Barbara K{\"o}nig, Stefan
  Milius, and Alexandra Silva.
\newblock A coalgebraic perspective on minimization and determinization.
\newblock In {\em International Conference on Foundations of Software Science
  and Computational Structures}, pages 58--73. Springer, 2012.
\newblock \href {https://doi.org/10.1007/978-3-642-28729-9_4}
  {\path{doi:10.1007/978-3-642-28729-9_4}}.

\bibitem{adamek2009abstract}
Jiri Adamek, Horst Herrlich, and George~E Strecker.
\newblock Abstract and concrete categories: The joy of cats.
\newblock {\em Reprints in Theory and Applications of Categories}, 2009.

\bibitem{adamek2019finitely}
Jiri Adamek, Stefan Milius, Lurdes Sousa, and Thorsten Wi{\ss}mann.
\newblock Finitely presentable algebras for finitary monads.
\newblock {\em Theory and Applications of Categories}, 34(37):1179--1195, 2019.

\bibitem{adamek1994locally}
Jiri Adamek and Jiri Rosicky.
\newblock {\em Locally Presentable and Accessible Categories}, volume 189.
\newblock Cambridge University Press, 1994.
\newblock \href {https://doi.org/10.1017/CBO9780511600579}
  {\path{doi:10.1017/CBO9780511600579}}.

\bibitem{angluin1987learning}
Dana Angluin.
\newblock Learning regular sets from queries and counterexamples.
\newblock {\em Information and Computation}, 75(2):87--106, 1987.
\newblock \href {https://doi.org/10.1016/0890-5401(87)90052-6}
  {\path{doi:10.1016/0890-5401(87)90052-6}}.

\bibitem{arbib1975fuzzy}
Michael~A Arbib and Ernest~G Manes.
\newblock Fuzzy machines in a category.
\newblock {\em Bulletin of the Australian Mathematical Society},
  13(2):169--210, 1975.
\newblock \href {https://doi.org/10.1017/S0004972700024412}
  {\path{doi:10.1017/S0004972700024412}}.

\bibitem{arnold1992note}
Andr{\'e} Arnold, Anne Dicky, and Maurice Nivat.
\newblock A note about minimal non-deterministic automata.
\newblock {\em Bulletin of the EATCS}, 47:166--169, 1992.

\bibitem{awodey2010category}
Steve Awodey.
\newblock {\em Category Theory}.
\newblock Oxford University Press, Inc., 2010.

\bibitem{beck1969distributive}
Jon Beck.
\newblock Distributive laws.
\newblock In {\em Seminar on Triples and Categorical Homology Theory}, pages
  119--140. Springer, 1969.
\newblock \href {https://doi.org/10.1007/BFb0083084}
  {\path{doi:10.1007/BFb0083084}}.

\bibitem{berndt2017learning}
Sebastian Berndt, Maciej Li{\'s}kiewicz, Matthias Lutter, and R{\"u}diger
  Reischuk.
\newblock Learning residual alternating automata.
\newblock In {\em Thirty-First AAAI Conference on Artificial Intelligence},
  2017.
\newblock \href {https://doi.org/10.1609/aaai.v31i1.10891}
  {\path{doi:10.1609/aaai.v31i1.10891}}.

\bibitem{bonchi2014algebra}
Filippo Bonchi, Marcello~M Bonsangue, Helle~H Hansen, Prakash Panangaden, Jan
  Rutten, and Alexandra Silva.
\newblock Algebra-coalgebra duality in brzozowski's minimization algorithm.
\newblock {\em ACM Transactions on Computational Logic (TOCL)}, 15(1):1--29,
  2014.
\newblock \href {https://doi.org/10.1145/2490818} {\path{doi:10.1145/2490818}}.

\bibitem{bonchi2012brzozowski}
Filippo Bonchi, Marcello~M Bonsangue, Jan Rutten, and Alexandra Silva.
\newblock Brzozowski’s algorithm (co)algebraically.
\newblock In {\em Logic and Program Semantics}, pages 12--23. Springer, 2012.
\newblock \href {https://doi.org/10.1007/978-3-642-29485-3_2}
  {\path{doi:10.1007/978-3-642-29485-3_2}}.

\bibitem{bousfield1977constructions}
Aldridge~K Bousfield.
\newblock Constructions of factorization systems in categories.
\newblock {\em Journal of Pure and Applied Algebra}, 9(2-3):207--220, 1977.
\newblock \href {https://doi.org/10.1016/0022-4049(77)90067-6}
  {\path{doi:10.1016/0022-4049(77)90067-6}}.

\bibitem{brzozowski1962canonical}
Janusz~A Brzozowski.
\newblock Canonical regular expressions and minimal state graphs for definite
  events.
\newblock In {\em Proc. Symposium of Mathematical Theory of Automata},
  volume~12, pages 529--561, 1962.

\bibitem{BrzozowskiT14}
Janusz~A. Brzozowski and Hellis Tamm.
\newblock Theory of {\'{a}}tomata.
\newblock {\em Theor. Comput. Sci.}, 539:13--27, 2014.
\newblock \href {https://doi.org/10.1016/j.tcs.2014.04.016}
  {\path{doi:10.1016/j.tcs.2014.04.016}}.

\bibitem{denis2001residual}
Fran{\c{c}}ois Denis, Aur{\'e}lien Lemay, and Alain Terlutte.
\newblock Residual finite state automata.
\newblock In {\em Annual Symposium on Theoretical Aspects of Computer Science},
  pages 144--157. Springer, 2001.
\newblock \href {https://doi.org/10.1007/3-540-44693-1_13}
  {\path{doi:10.1007/3-540-44693-1_13}}.

\bibitem{esposito2002learning}
Yann Esposito, Aur{\'e}lien Lemay, Fran{\c{c}}ois Denis, and Pierre Dupont.
\newblock Learning probabilistic residual finite state automata.
\newblock In {\em International Colloquium on Grammatical Inference}, pages
  77--91. Springer, 2002.
\newblock \href {https://doi.org/10.1007/3-540-45790-9_7}
  {\path{doi:10.1007/3-540-45790-9_7}}.

\bibitem{jacobs2006bialgebraic}
Bart Jacobs.
\newblock A bialgebraic review of deterministic automata, regular expressions
  and languages.
\newblock In {\em Algebra, Meaning, and Computation}, pages 375--404. Springer,
  2006.
\newblock \href {https://doi.org/10.1007/11780274_20}
  {\path{doi:10.1007/11780274_20}}.

\bibitem{jacobs2011bases}
Bart Jacobs.
\newblock Bases as coalgebras.
\newblock In {\em Algebra and Coalgebra in Computer Science}, pages 237--252.
  Springer, 2011.
\newblock \href {https://doi.org/10.1007/978-3-642-22944-2_17}
  {\path{doi:10.1007/978-3-642-22944-2_17}}.

\bibitem{jacobs2017introduction}
Bart Jacobs.
\newblock {\em Introduction to Coalgebra: Towards Mathematics of States and
  Observation}.
\newblock Cambridge Tracts in Theoretical Computer Science. Cambridge
  University Press, 2016.
\newblock \href {https://doi.org/10.1017/CBO9781316823187}
  {\path{doi:10.1017/CBO9781316823187}}.

\bibitem{jacobs2012trace}
Bart Jacobs, Alexandra Silva, and Ana Sokolova.
\newblock Trace semantics via determinization.
\newblock In {\em International Workshop on Coalgebraic Methods in Computer
  Science}, pages 109--129. Springer, 2012.
\newblock \href {https://doi.org/10.1007/978-3-642-32784-1_7}
  {\path{doi:10.1007/978-3-642-32784-1_7}}.

\bibitem{kurzlogics}
Alexander Kurz.
\newblock {\em Logics for Coalgebras and Applications to Computer Science}.
\newblock PhD thesis, Ludwig-Maximilians-Universität München, 2000.

\bibitem{lang2004algebra}
Serge Lang.
\newblock Algebra.
\newblock {\em Graduate Texts in Mathematics}, 2002.

\bibitem{mac2013categories}
Saunders Mac~Lane.
\newblock {\em Categories for the Working Mathematician}, volume~5.
\newblock Springer, 2013.
\newblock \href {https://doi.org/10.1007/978-1-4757-4721-8}
  {\path{doi:10.1007/978-1-4757-4721-8}}.

\bibitem{maclane1950duality}
Saunders MacLane.
\newblock Duality for groups.
\newblock {\em Bulletin of the American Mathematical Society}, 56(6):485--516,
  1950.

\bibitem{moggi1988computational}
Eugenio Moggi.
\newblock {\em Computational Lambda-Calculus and Monads}.
\newblock University of Edinburgh, Department of Computer Science, Laboratory
  for Foundations of Computer Science, 1988.

\bibitem{moggi1990abstract}
Eugenio Moggi.
\newblock {\em An Abstract View of Programming Languages}.
\newblock University of Edinburgh, Department of Computer Science, Laboratory
  for Foundations of Computer Science, 1990.

\bibitem{moggi1991notions}
Eugenio Moggi.
\newblock Notions of computation and monads.
\newblock {\em Information and Computation}, 93(1):55--92, 1991.
\newblock \href {https://doi.org/10.1016/0890-5401(91)90052-4}
  {\path{doi:10.1016/0890-5401(91)90052-4}}.

\bibitem{MyersAMU15}
Robert S.~R. Myers, Jiri Adamek, Stefan Milius, and Henning Urbat.
\newblock Coalgebraic constructions of canonical nondeterministic automata.
\newblock {\em Theoretical Computer Science}, 604:81--101, 2015.
\newblock \href {https://doi.org/10.1016/j.tcs.2015.03.035}
  {\path{doi:10.1016/j.tcs.2015.03.035}}.

\bibitem{nerode1958linear}
Anil Nerode.
\newblock Linear automaton transformations.
\newblock {\em Proceedings of the American Mathematical Society},
  9(4):541--544, 1958.
\newblock \href {https://doi.org/10.2307/2033204} {\path{doi:10.2307/2033204}}.

\bibitem{parlant_et_al:LIPIcs:2020:12746}
Louis Parlant, Jurriaan Rot, Alexandra Silva, and Bas Westerbaan.
\newblock {Preservation of Equations by Monoidal Monads}.
\newblock In {\em 45th International Symposium on Mathematical Foundations of
  Computer Science (MFCS 2020)}, volume 170 of {\em Leibniz International
  Proceedings in Informatics (LIPIcs)}, pages 77:1--77:14. Schloss
  Dagstuhl--Leibniz-Zentrum f{\"u}r Informatik, 2020.
\newblock \href {https://doi.org/10.4230/LIPIcs.MFCS.2020.77}
  {\path{doi:10.4230/LIPIcs.MFCS.2020.77}}.

\bibitem{riehl2008factorization}
Emily Riehl.
\newblock Factorization systems.
\newblock 2008.
\newblock URL: \url{https://math.jhu.edu/~eriehl/factorization.pdf}.

\bibitem{rutten2000universal}
Jan Rutten.
\newblock Universal coalgebra: A theory of systems.
\newblock {\em Theoretical Computer Science}, 249(1):3--80, 2000.
\newblock \href {https://doi.org/10.1016/S0304-3975(00)00056-6}
  {\path{doi:10.1016/S0304-3975(00)00056-6}}.

\bibitem{rutten2019method}
Jan Rutten.
\newblock The method of coalgebra: Exercises in coinduction.
\newblock 2019.

\bibitem{seal2013tensors}
Gavin~J Seal.
\newblock Tensors, monads and actions.
\newblock {\em Theory and Applications of Categories}, 28(15):403--433, 2013.

\bibitem{silva2010generalizing}
Alexandra Silva, Filippo Bonchi, Marcello~M Bonsangue, and Jan Rutten.
\newblock Generalizing the powerset construction, coalgebraically.
\newblock In {\em IARCS Annual Conference on Foundations of Software Technology
  and Theoretical Computer Science (FSTTCS 2010)}, volume~8, pages 272--283.
  Schloss Dagstuhl -- Leibniz-Zentrum fuer Informatik, 2010.
\newblock \href {https://doi.org/10.4230/LIPIcs.FSTTCS.2010.272}
  {\path{doi:10.4230/LIPIcs.FSTTCS.2010.272}}.

\bibitem{STREET1972149}
Ross Street.
\newblock The formal theory of monads.
\newblock {\em Journal of Pure and Applied Algebra}, 2(2):149--168, 1972.
\newblock \href {https://doi.org/https://doi.org/10.1016/0022-4049(72)90019-9}
  {\path{doi:https://doi.org/10.1016/0022-4049(72)90019-9}}.

\bibitem{Street2009}
Ross Street.
\newblock Weak distributive laws.
\newblock {\em Theory and Applications of Categories}, 22:313--320, 2009.

\bibitem{turi1996functorial}
Daniele Turi.
\newblock {\em Functorial Operational Semantics}.
\newblock PhD thesis, Vrije Universiteit Amsterdam, 1996.

\bibitem{turi1997towards}
Daniele Turi and Gordon Plotkin.
\newblock Towards a mathematical operational semantics.
\newblock In {\em Proceedings of Twelfth Annual IEEE Symposium on Logic in
  Computer Science}, pages 280--291. IEEE, 1997.
\newblock \href {https://doi.org/10.1109/LICS.1997.614955}
  {\path{doi:10.1109/LICS.1997.614955}}.

\bibitem{van2016master}
Gerco van Heerdt.
\newblock An abstract automata learning framework.
\newblock Master's thesis, Radboud University Nijmegen, 2016.

\bibitem{van2020phd}
Gerco van Heerdt.
\newblock {\em {CALF}: Categorical Automata Learning Framework}.
\newblock PhD thesis, University College London, 2020.

\bibitem{van2020learning}
Gerco van Heerdt, Matteo Sammartino, and Alexandra Silva.
\newblock Learning automata with side-effects.
\newblock In {\em Coalgebraic Methods in Computer Science}, pages 68--89.
  Springer, 2020.
\newblock \href {https://doi.org/10.1007/978-3-030-57201-3_5}
  {\path{doi:10.1007/978-3-030-57201-3_5}}.

\bibitem{VuilleminG210}
Jean Vuillemin and Nicolas Gama.
\newblock Efficient equivalence and minimization for non deterministic xor
  automata.
\newblock Technical report, {Ecole Normale Sup{\'e}rieure}, 2010.

\bibitem{wissmann2022minimality}
Thorsten Wißmann.
\newblock Minimality notions via factorization systems and examples.
\newblock {\em Logical Methods in Computer Science}, 18(3), 2022.
\newblock \href {https://doi.org/10.46298/lmcs-18(3:31)2022}
  {\path{doi:10.46298/lmcs-18(3:31)2022}}.

\bibitem{zetzsche2021}
Stefan Zetzsche, Gerco van Heerdt, Matteo Sammartino, and Alexandra Silva.
\newblock Canonical automata via distributive law homomorphisms.
\newblock {\em Electronic Proceedings in Theoretical Computer Science},
  351:296–313, 2021.
\newblock \href {https://doi.org/10.4204/eptcs.351.18}
  {\path{doi:10.4204/eptcs.351.18}}.

\end{thebibliography}

\end{document}